# Optical coherent dot-product chip for sophisticated deep learning regression

Shaofu Xu (xsf19950411@sjtu.edu.cn)[1],
Jing Wang (wj1996@sjtu.edu.cn)[1],
Haowen Shu (haowenshu@pku.edu.cn)[2],
Zhike Zhang (zkzhang@semi.ac.cn)[3],
Sicheng Yi (707197ysc@sjtu.edu.cn)[1],
Bowen Bai (bowenbai@pku.edu.cn)[2],
Xingjun Wang (xjwang@pku.edu.cn)[2],
Jianguo Liu (jgliu@semi.ac.cn)[3],
Weiwen Zou (wzou@sjtu.edu.cn)[1]*

*[1]State Key Laboratory of Advanced Optical Communication Systems and Networks, Intelligent Microwave Lightwave Integration Innovation Center (imLic), Department of Electronic Engineering, Shanghai Jiao Tong University, 800 Dongchuan Road, Shanghai 200240, China*
*[2]State Key Laboratory of Advanced Optical Communications System and Networks, Department of Electronics, School of Electronics engineering and Computer Science, Peking University, Beijing, 100871, China*
*[3]Institution of Semiconductors, Chinese Academy of Sciences, Beijing 100083, China*
**Correspondence to: Weiwen Zou (Email: wzou@sjtu.edu.cn, Tel: +86-13917928669).*

***Abstract:*** Optical implementations of neural networks (ONNs) herald the next-generation high-speed and energy-efficient deep learning computing by harnessing the technical advantages of large bandwidth and high parallelism of optics. However, due to the problems of incomplete numerical domain, limited hardware scale, or inadequate numerical accuracy, the majority of existing ONNs were studied for basic classification tasks. Given that regression is a fundamental form of deep learning and accounts for a large part of current artificial intelligence applications, it is necessary to master deep learning regression for further development and deployment of ONNs. Here, we demonstrate a silicon-based optical coherent dot-product chip (OCDC) capable of completing deep learning regression tasks. The OCDC adopts optical fields to carry out operations in complete real-value domain instead of in only positive domain. Via reusing, a single chip conducts matrix multiplications and convolutions in neural networks of any complexity. Also, hardware deviations are compensated via *in-situ* backpropagation control provided the simplicity of chip architecture. Therefore, the OCDC meets the requirements for sophisticated regression tasks and we successfully demonstrate a representative neural network, the AUTOMAP (a cutting-edge neural network model for image reconstruction). The quality of reconstructed images by the OCDC and a 32-bit digital computer is comparable. To the best of our knowledge, there is no precedent of performing such state-of-the-art regression tasks on ONN chip. It is anticipated that the OCDC can promote novel accomplishment of ONNs in modern AI applications including autonomous driving, natural language processing, and scientific study.

## Introduction

Because of the flourishment of artificial intelligence (AI), we witness the revolution of technical foundations of emerging applications, such as autonomous driving, natural language processing, and medical diagnosis [1-3]. Additionally, profound insights are offered to scientific studies across disciplines such as chemistry [4], physics [5, 6], and biomedicine [7]. One of the major driving forces of AI is the blooming of artificial neural networks (ANNs), which are mathematically composed of thousands of nodes and millions of interconnections layer by layer. A high-dimension representation space is thus supported by the large-scale neural network. Such large representation space of ANNs enables high-volume automatic feature extraction from original data, so that intricate transformations of chemical, physical, and biological systems can be precisely fitted and predicted. However, large representation space demands massive computational cost. Currently, the Moore's law of integrated circuits is slowing down [8] while the scale expansion of ANNs is speeding up [9]. The compute capability of conventional digital computers is falling behind. To solve the problem, optical implementations of neural network (ONNs) have been recently proposed and demonstrated to realize high-speed and energy-efficient AI hardware [10-14]. Linear propagation of light equivalently carries out the linear computing of ANNs [15-19]; ultra-wide optical transparent spectrum and high-speed modulators/detectors enable a fast clock rate (tens of GHz) [20, 21]; and non-volatile photonic memory makes the computing 'zero-consuming' [22].

However, a substantial improvement of ONNs needs to be achieved for accomplishing state-of-the-art AI applications. For now, ONNs are mostly demonstrated with classification tasks on elementary datasets such as MNIST handwritten digit recognition [23] because of their simplicity for primary validation. As an important form of deep learning, regression tasks, such as image reconstruction [24], machine gaming [25], and nanostructure design [26, 27], remain uninvestigated. Distinct from classification, regression demands the neural network to output continuous values instead of discrete categories. Carrying out computations in the complete real-value domain with high numerical accuracy is the basic requirement for regression, which is still challenging for the existing ONN chips. Firstly, for non-coherent ONN architectures [16, 20-22], input values are represented by non-negative optical intensities, causing incompleteness of the numerical domain. In contrast, coherent ONN architectures [15, 28, 29] adopt optical fields to represent real-valued inputs and homodyne detection to yield real-valued outputs, showing the capability of computing in the complex-value domain. Nonetheless, the size of existing ONN chips is much smaller than that of regression neural networks and the complexity of chip calibration for coherent ONNs increases the difficulty of reaching high numerical accuracy. Therefore, high-quality deep learning regression still remains challenging in the ONN field.

Here, we propose and experimentally demonstrate a silicon-based optical coherent dot-product chip (OCDC) to implement sophisticated regression tasks. Values are modulated into the amplitudes of optical fields and the output field is read out via optical interference. In this sense, it is feasible to operate the OCDC in the complete real-value domain. The chip is reconfigured to conduct linear operations including matrix multiplications and convolutions, and it is reused to carry out arbitrarily sophisticated neural networks. As the OCDC is an analog computing device, parameters represented by nonideal hardware often deviate from desired ones. The simple architecture of the OCDC enables compensation for such deviations by *in-situ* backpropagation control, thus obviously enhancing numerical accuracy. With these properties, the chip meets the basic prerequisites for deep learning regression tasks. We benchmark the OCDC with a neural network, AUTOMAP [30], which achieves state-of-the-art performance in image reconstruction. Experimental result verifies that the OCDC can effectively compute both fully-connected layers and convolutional layers, covering all linear operations required by most neural networks. The performance of OCDC in the AUTOMAP image reconstruction task is comparable with that of a 32-bit digital computer. To the best of our knowledge, it is the first time of demonstrating sophisticated deep learning regression tasks with on-chip optical computing hardware. The insights provided by this work is inspiring for further investigations on practically applicable ONNs.

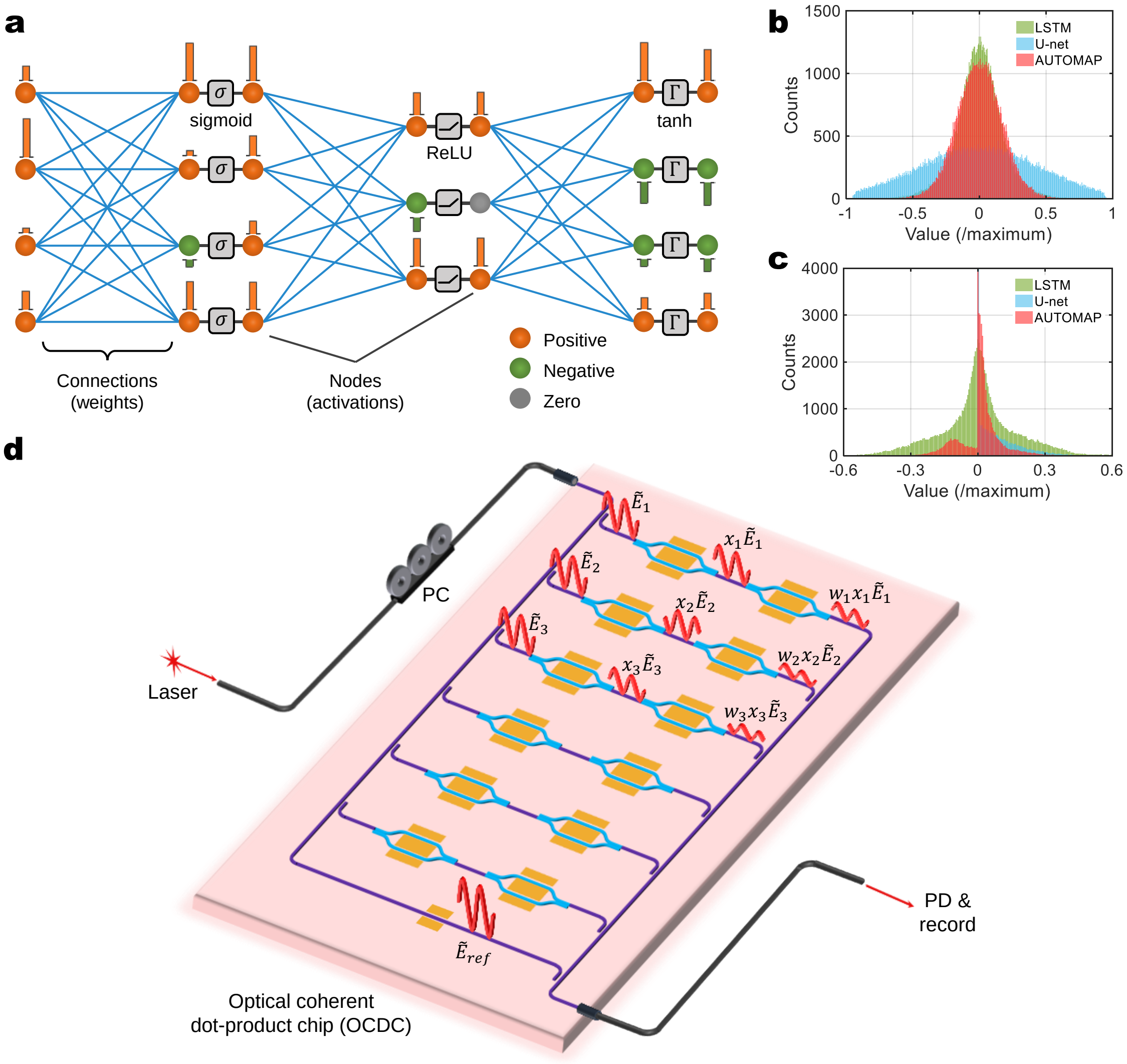

**Fig.1** Schematic of optical coherent dot-product chip (OCDC). **a**, Simplified model of a neural network for regression. A neural network is typically composed with nodes, connections, and activation functions. Three typical functions (sigmoid, rectified linear unit (ReLU), and hyperbolic tangent (tanh)) are depicted in the figure. Positive, negative and zeros nodes are depicted with different colors. For fully-connected networks, a node represents a single value and a connection is a weight. For convolutional networks, a node is a feature map and a connection is a convolutional kernel. **b**, Histograms of weights in LSTM [31], U-net [32] and AUTOMAP [30], respectively. They approximately obey normal distribution with mean value of zero. **c**, Histograms of activated values in LSTM, U-net and AUTOMAP, respectively. The distribution depends on the activation functions being used. **d**, The conceptual schematic of the OCDC. It contains several parallel branches for dot product and one extra branch for coherent detection. The optical field in each branch is symbolized with red curves. The push-pull configured modulator imposes amplitude-only modulation to the optical field without introducing phase shift. Hence, the phase of each branch is stable.

In Fig. 1a, a simplified model of an ANN for regression tasks is depicted. The network is composed with nodes and connections layer by layer. A node represents a single value in fully-connected (FC) layers or a feature map in convolutional (conv.) layers. Connections represent the weight matrix or convolutional kernels for FC layer or conv. layer correspondingly. Each layer contains a linear part, i.e., matrix-vector multiplication (MVM) and convolution, and a nonlinear activation function to obtain the activated values. From left to right, the input nodes are calculated layer by layer to yield the final output nodes. For regression tasks, weights in the network are trained to minimize the distance between the final outputs and the ground truth (data deemed as reference). In general cases of regression networks, the numerical basis for both activated values and weights is the real-value domain. In Fig. 1b and 1c, we illustrate the histograms of weights and activated values of three well-known regression ANNs, U-net [32], long short-term memory (LSTM) [31], and AUTOMAP [30]. The overall distribution of trained weights in these ANNs always obey normal distributions and the mean values are near zero. Negative weights are as many as positive ones. The activated values perform diversely. In the U-net with ReLU activation function, activated values are non-negative. For this kind of neural networks, only positive input values are need and non-coherent ONN architecture performs similarly with coherent architectures. However, the LSTM network and the AUTOMAP, which are based on hybrid sigmoid, ReLU, and tanh functions, contain both positive and negative activated values. Therefore, the capability of representing real activated values and weights is a basic requirement for optical implementation of regression networks. Also, we note that the dot product is a building block of MVMs or convolutions. Building a hardware for dot product enables equivalent calculation of the linear part of ANNs. Following these guidelines, we design the OCDC as shown in Fig. 1d. To keep the signs of activated values and weights, optical amplitude modulation is adopted and the output amplitude is detected via coherent interference. The light from a coherent laser source is split into M+1 branches, with M branches performing the dot product and the last one being the local reference for coherent detection. Optical power is evenly distributed in these M branches. Inside each branch, two modulators under push-pull configuration are deployed serially to impose optical amplitude modulation without introducing extra phase shift. As the first one represents the value of $x_i$ and the next one represents $w_i$, the output is the multiplication of these two values [18]. When all branches match in phase, the optical fields interfere constructively in the optical combiner to complete summation. Before photodetection, the local reference is combined with the summed optical field, introducing an amplitude bias to avoid elimination of negative amplitude at photodetection. The process of the OCDC can be formulated as the following equation.

$$
\begin{aligned}
I_{photo} &\propto \mathbf{Re}\left\{\left(\tilde{E}_{ref}+\sum_i w_i x_i \tilde{E}_i\right)\left(\tilde{E}_{ref}+\sum_i w_i x_i \tilde{E}_i\right)^*\right\} \\
&= \left\|\tilde{E}_{ref}+\sum_i w_i x_i \tilde{E}_i\right\|^2
\end{aligned}
\qquad (1)
$$

When the amplitude of local reference $\tilde{E}_{ref}$ is larger than the weighted sum, the sign of dot product is maintained after photodetection.

## Results

The OCDC is fabricated with a silicon-on-insulator (SOI) process. Figure 2a shows the packaged OCDC and its periphery circuits. Common metal wires on the bottom provides voltages for the chip. Transmission lines on the top are used to transfer high-speed signals. The layout of the chip is shown in Fig. 2b. Optical splitters, push-pull modulators, combiners, couplers are systematically integrated on the chip. Two optical splitters (shown in Fig. 2c) are used to divide optical power into the nine modulating branches and the reference. A multi-mode interferometer (MMI) split a half of optical power to the reference. The left power is split into 9 branches by cascaded directional couplers (DCs). The coupling length of each DC is designed so that the optical power is divided evenly. Figure 2d provides the measured splitting ratio of the cascaded DCs, showing an evenness lower than 1.2 dB. At each modulating branch, we fabricate a tail phase shifter to compensate for the phase difference between branches and the reference. More details of chip fabrication and characterization are provided in the Suppl. In the OCDC, stable push-pull modulation is important to keep the constructive interference for photodetection. Figure 2e shows an example result of push-pull modulation (see Methods). This result is yielded by complementarily changing voltages on the upper and lower arms of a single modulator with other modulators stay static. The amplitude of output optical field varies along a cosine curve when the applied voltages change. The $R^2$ of fitting is 0.9994, indicating that the push-pull modulation is accomplished with high stability. Then the constructive interference of multiple branches is inspected in Fig. 2f. It is shown that, with proper configuration of the tail phase shifter, the output optical fields of different branches can match in-phase, performing a jointly constructive interference. Implied by the results, the OCDC is capable of amplitude modulation and coherent detection, laying the basis for calculations of real-valued dot product.

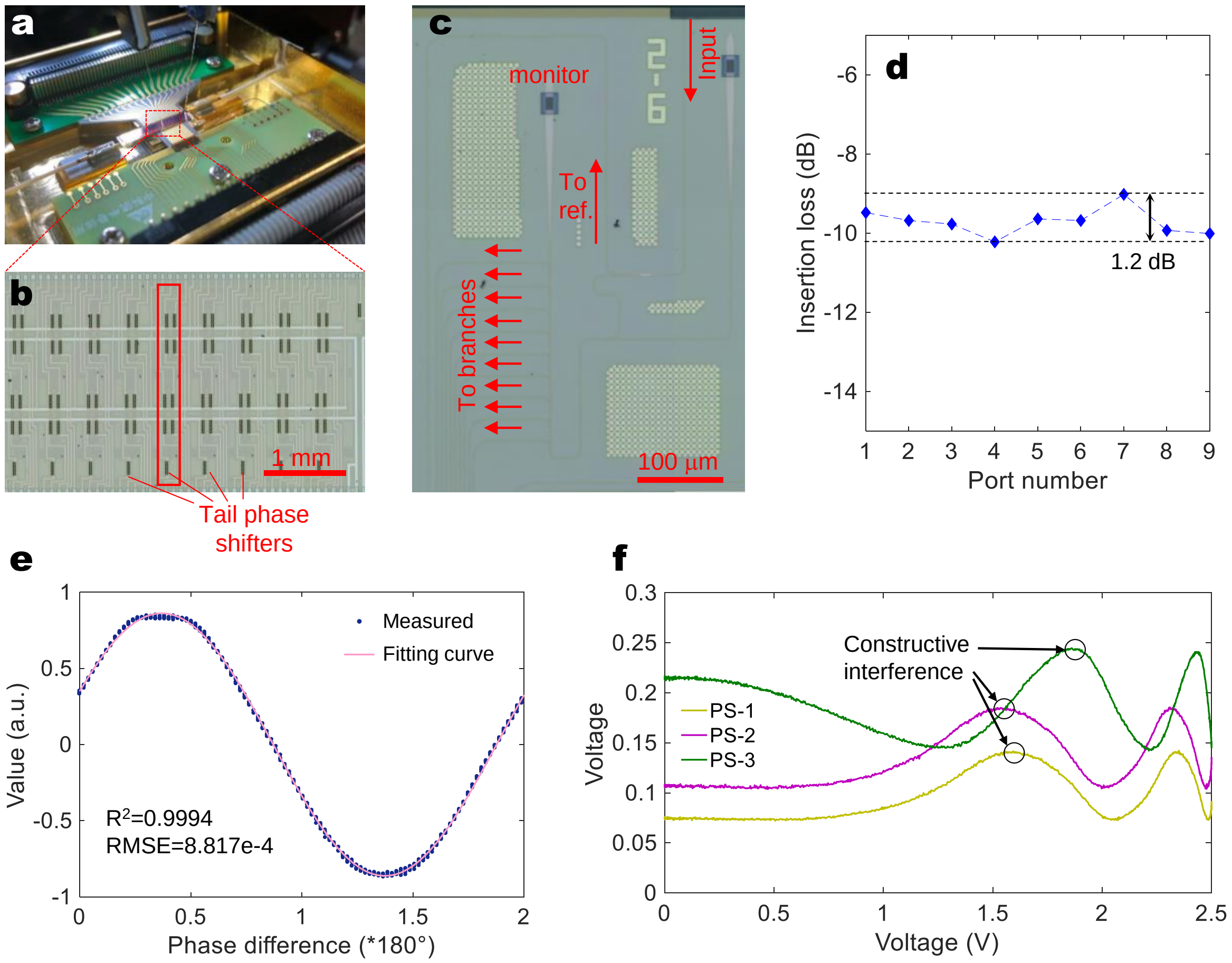

**Fig. 2** Chip characterization results. **a**, The packaged OCDC with periphery circuits. **b**, The top view of the OCDC. Modulating branches are located vertically as the red block shows. Every modulator contains four phase shifters, two of which are used to conduct push-pull modulation and the remaining two are used to control the bias voltage. Tail phase shifters are appended to compensate for the phase difference among branches. **c**, The structure of light input and splitter. **d**, Characterization of evenness of optical power splitting. **e**, Modulator characterization with push-pull driven. The experimental result is fitted with a curve formulated by *a*sin(b*x+c)+d*. The information of fitting is also given in the plot. **f**, The effect of constructive interference of three branches. Three curves are obtained one by one. Firstly, only the first branch is modulated to the highest transparency when other two branches are closed. By changing the voltage on its tail phase shifter (PS-1), the yellow curve is shown. It denotes the interference result of branch 1 and reference branch. Secondly, keep the voltage on the PS-1 at the constructive interference point (black circled); modulate the second branch to its highest transparency; change the second tail phase shifter (PS-2). The purple curve is recorded. By operating the same process for the third branch, we obtain the green curve.

We implement the AUTOMAP as a representative example of deep learning regression to validate the OCDC. Figure 3a shows the structure of the AUTOMAP containing two FC layers, two convolutional layers, and a de-convolutional layer. Details of the neural network can be found at ref. [30] and Methods. Given that the linear part of FC layers and convolutional layers can be decomposed to dot products; we can realize these layers by reusing the OCDC temporally. Fig. 3b-d show the method of conducting MVMs and convolutions via temporal multiplexing. The decomposition of MVMs is straightforward since they are naturally calculated via vector-vector dot product. The input vector is loaded onto the second row of modulators marked ‘slow mod.’, and a vector from the matrix is modulated onto the ‘fast mod.’ modulators. By temporally changing the vectors loaded on the ‘fast mod.’ modulators, the result of MVM is eventually calculated. When the size of vectors is too large to be loaded onto these modulators in one time, the vector and the matrix can be divided into small parts as depicted in Fig. 3c. For convolutions, the process of patching [33, 34] can rearrange pixels of the feature map into a matrix. The kernel is flattened as a vector. In this way, MVMs and convolutions can be similarly conducted by the OCDC. Fig. 3d is the experimental setup (detailed in Methods) for temporally multiplexing the OCDC. A signal generator (max. bandwidth is 20 MHz) is used to provide signals for amplitude modulation and a voltage source (VS) supplies the bias voltages. A computer (the grey block) is adopted to carry out programs. It controls the signal generator and the VS to work as a whole. It also records and processes the output data from the OCDC.

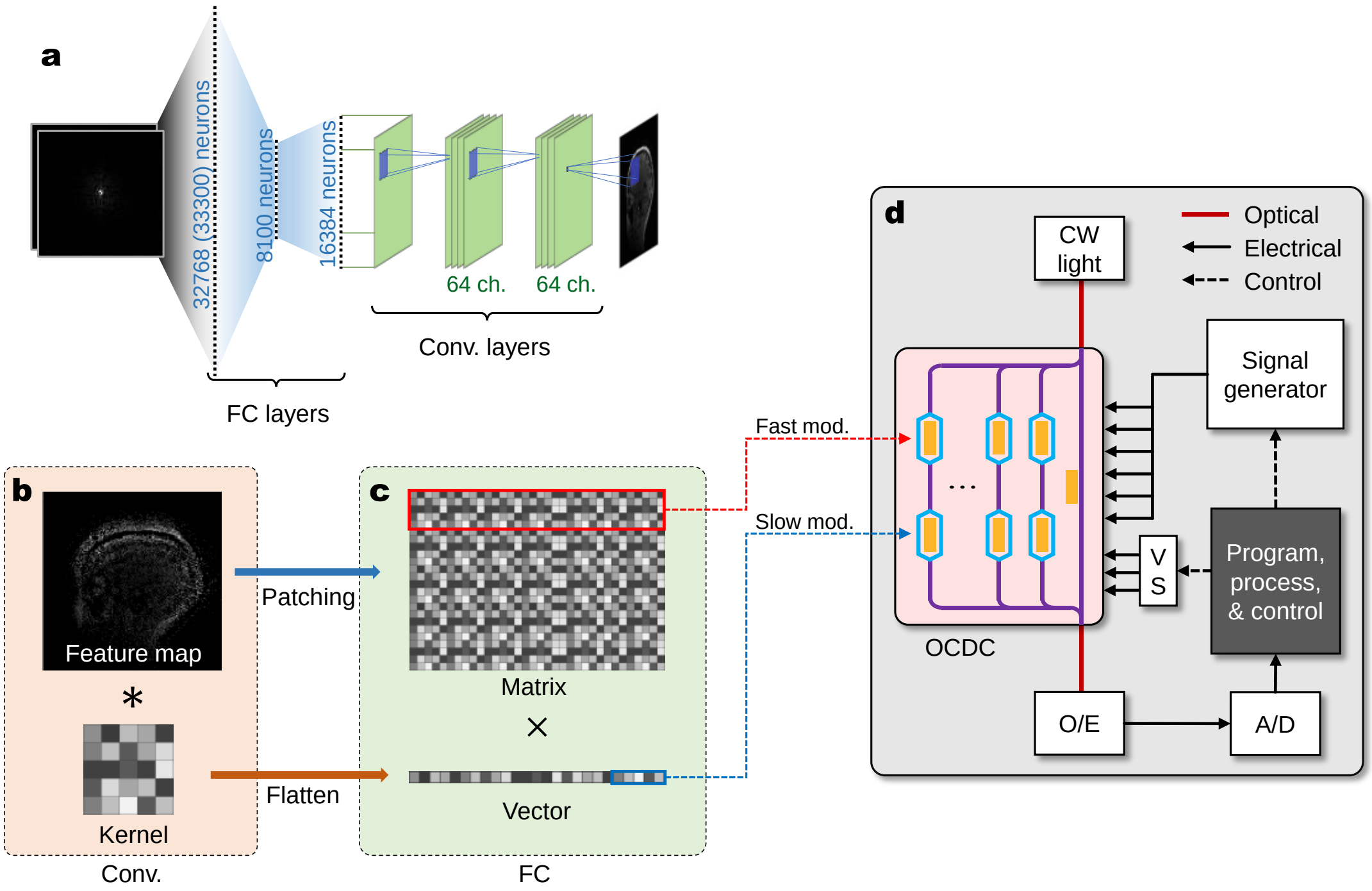

**Fig. 3** Experimental setup of OCDC validation. **a**, The network structure of the AUTOMAP. The input images are flattened into vectors and process with two fully-connected layers. The numbers of neurons are 32768 (or 33300), 8100, and 16384, respectively. Then, the output neurons are converted back to feature maps sized 128×128. Two convolutional layers and a de-convolutional layer process this feature map to the output image. The number of channels in convolutional layers is 64. **b**, The process of convolutions. After patching and flattening, convolutions can be converted to MVMs. **c**, The process of MVM. The matrix and the vector are loaded onto the fast mod and slow mod., respectively. The red and blue blocks show a decomposition of the input vector and matrix. **d**, Experimental setup of OCDC validation. CW light, continuous-wave light source; O/E, optoelectronic detection; A/D, analog-to-digital converter.

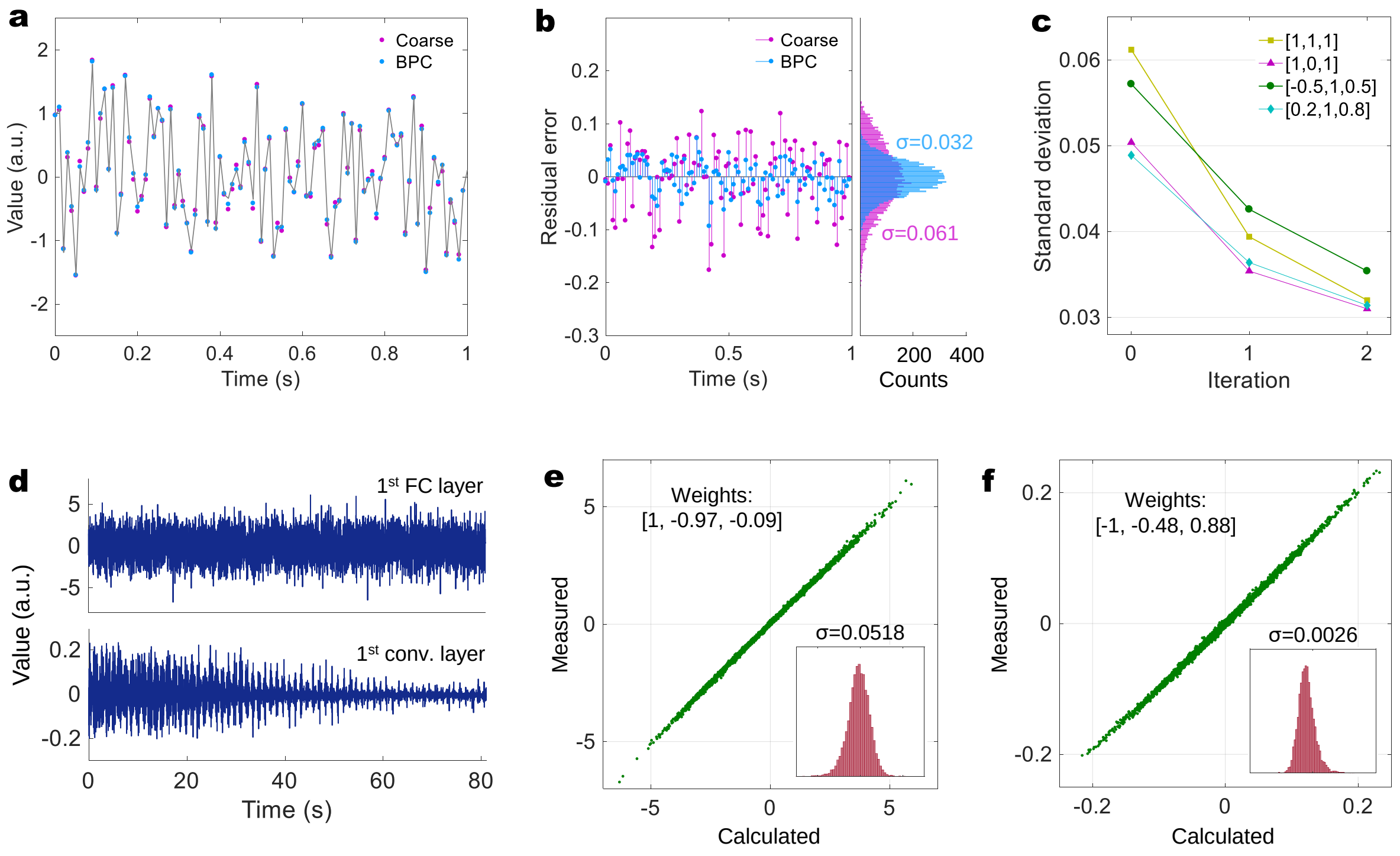

**Fig. 4** Experimental results of OCDC validation. **a**, Samples before and after the BPC. The 'coarse' samples are yielded with the coarsely-calibrated hardware. The weights adopted is [1, 1, 1]. The grey curve shows the correct result calculated by a computer. **b**, Residual error of the samples before and after the BPC process. Histograms and standard deviations (σ) are also attached on the right side. **c**, BPC process with different weight combination. **d**, Output waveforms of the OCDC conducting the 1$^{st}$ FC layer and the 1$^{st}$ convolutional layer, respectively. **e, f**, Scatter plots of the measured samples vs. calculated expectation (FC layer and conv. layer, respectively). Corresponding weights are marked on the top-left. Histogram of residual error is attached on the bottom-right.

As an analog computing hardware, the OCDC suffers from the imperfectness of fabricated devices. The actual values represented by the analog devices often deviate from the desired ones. Such deviations come from multiples sources including uneven splitter, combiner, modulation efficiency, and phase drift. Distinct from classification tasks, deep learning regression asks for higher numerical accuracy since it directly corresponds to the quality of regression (See Fig. S9 for more information). Therefore, we adopt an *in-situ* backpropagation control (BPC) method to minimize such deviations (see Methods). Figures 4a and 4b show the effectiveness of the BPC with an example of random inputs. The weights adopted is [1, 1, 1]. Although the modulators are coarsely calibrated before conducting dot products, the numerical accuracy is insufficient for high-quality deep learning regression. After the BPC, the accuracy of the analog computing is improved with residual error dropping from 0.061 to 0.032. With more weight combinations (see Fig. 4c and Fig. S10), we validate that the BPC can increase the numerical accuracy of the coarsely-calibrated hardware within 2 iterations.

Based on the BPC method, we experimentally implement the first FC layer and the first convolutional layer of AUTOMAP using the OCDC. Due to the limited hardware scale, the parameters are trained on a computer and the OCDC is used for inference (see the Methods for details of network training). For the FC layer, the sizes of input vector (1×32768) and the weight matrix (32768×8100) are massive. We decompose the vector and the matrix to the size of 1×3 and 3×8100, respectively, according to the size of the OCDC. For the convolutional layer, the input feature map (128×128) is firstly patched to a matrix (25×16384) and the kernel (5×5) is flattened. Similarly, we decompose the massive matrix to small parts for feasibility of the OCDC. The OCDC carries out the linear parts and the nonlinear activation functions are implemented in the computer. Further details of experiment are provided in Methods. An example result is shown in Fig. 4d. Each temporal waveform contains 8100 samples since the OCDC is temporally multiplexed by 8100 times. Limited by the bandwidth of thermo-optic modulator and the signal generator, the modulation rate is 100 Hz. A high-speed OCDC is further discussed in Suppl. In Fig. 4e and 4f, the computing accuracy of the FC layer and the convolutional layer is inspected, respectively. We observe that the measured samples tightly concentrate at the diagonal line where indicates correct results. The normalized standard deviation of the residual error is 0.0518/7.0=0.0074 and 0.0026/0.25=0.0104, respectively. Such results are competitive among coherent ONN architectures, around six times less than that reported in ref. [15].

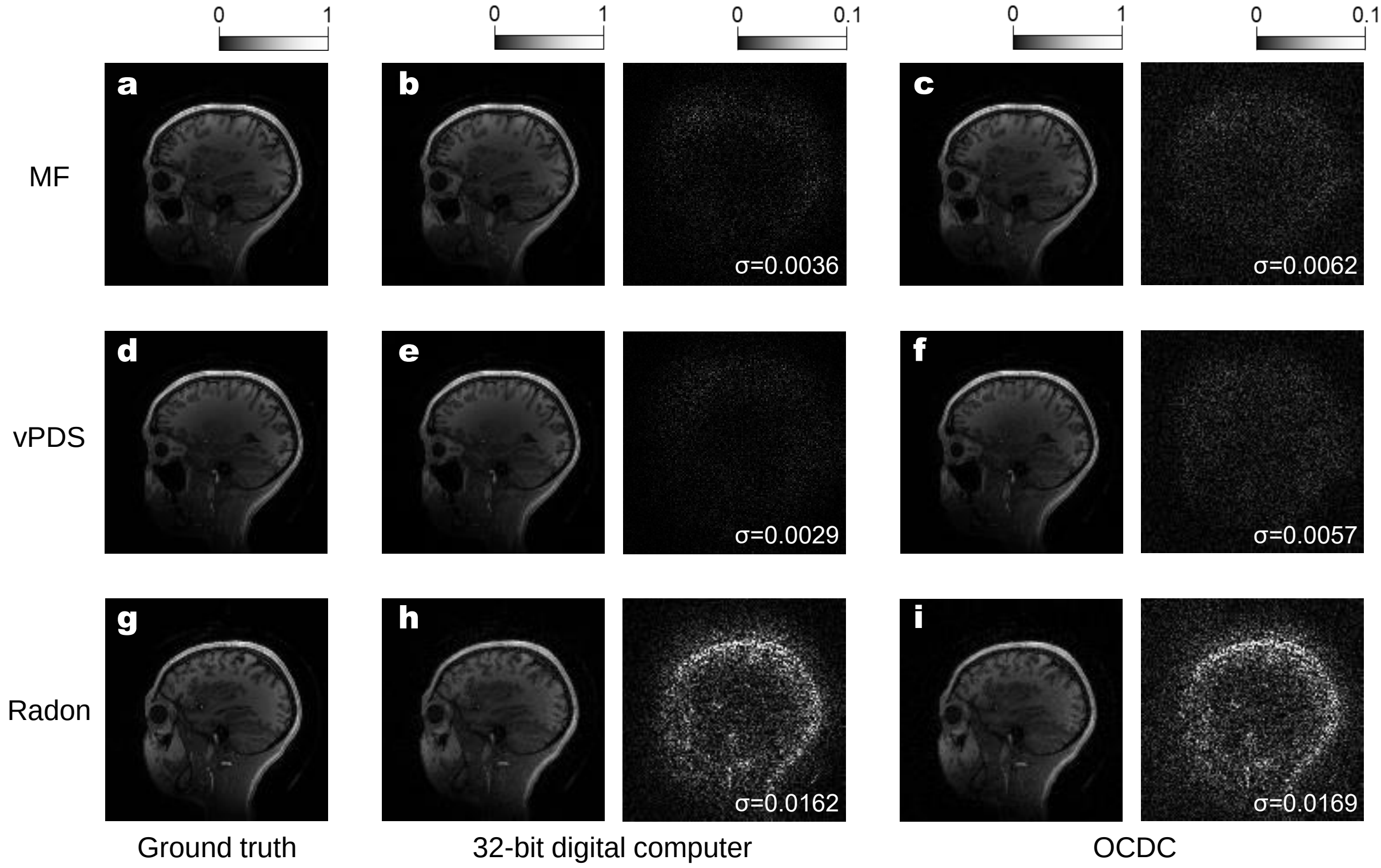

**Fig. 5** Reconstructed images of the AUTOMAP by a 32-bit computer and the OCDC. **a, d, g,** The ground-truth images with MF, vPDS, and Radon processes, respectively. **b, e, h,** The reconstructed images by the 32-bit computer. Values of the images are normalized to 1. Scale bars are attached on top. The residual error maps are attached on the right. Values of the error are amplified by 10 times for better visibility. Standard deviations (σ) are shown on the error maps. **c, f, i,** Reconstructed images of the OCDC, with the same normalization.

Then, the task of image reconstruction is demonstrated with the accuracy achieved by the OCDC (details are provided in Methods). Example results are shown in Fig. 5 and Figs. S12-14. Since the AUTOMAP is a unified neural network which can reconstruct images from various input formats [30], we demonstrate three typical reconstruction processes for magnet resonance imaging (MRI): misaligned Fourier (MF) space [35], variable Poisson disk sampled (vPDS) Fourier space [36], and Radon projection [37] (see Method for further details). Figs. 5a-5c illustrate a comparison of the results yielded by a standard 32-bit computer and the OCDC. The corresponding process is MF. We observe that the OCDC accomplishes image reconstruction with high quality. The standard deviations of image absolute error are 0.0036 and 0.0062 for the digital computer and the OCDC, respectively. Note that the values of error are amplified by 10 times for better visibility. Such increase of error of the OCDC is hardly visible from the reconstructed image. Figures 5d-5f are the reconstructed images of the vPDS process. The quality of the OCDC reconstructed image is also acceptable. For the Radon reconstruction, performance of the AUTOMAP conducted by the computer is inferior to the previous two processes. The gap between the 32-bit computer and the OCDC is greatly reduced. From the results, we observe that the OCDC can achieve a comparable performance with 32-bit digital computer on image reconstruction, implying its further applications on other regression tasks.

## Discussion

Researches on ONNs are commonly in pursuit of high-speed and energy-efficient computing. Although the main contribution of this work is originally demonstrating regression tasks, we note that the proposed architecture is potential to be upgraded towards high speed one. Recent breakthroughs of electro-optic integrated modulators, on conventional SOI platform or thin-film lithium niobate platform [38-40], pave the way for high modulation rate. Replacing the thermo-optic modulators used in the proof-of-concept (the 'fast mod.' part) can greatly enhance the computing speed. In supplementary material, an electro-optic version of OCDC is measured and discussed. If the OCDC is designed at high speed, the system noise will increase inevitably. The signal-to-noise ratio (SNR) might become a limiting factor to the performance of the OCDC. Generally speaking, SNR is mostly determined by the insertion loss and photodetection noise [12]. If the insertion loss and the photodetection noise are kept low with current advanced waveguide and photodetection technologies, high SNR is achievable at high modulation rate over 10 GHz and high-quality image reconstruction is also expectable. The OCDC has the potential to be duplicated spatially for exploiting the optical advantage on parallelism (described detailly in Suppl.). The operation speed can thus be multiplied and the energy consumption per operation can be lowered [11-13, 19]. Moreover, the phase shifters in the 'slow mod.' part can be replaced with non-thermal devices such as nano optical electromechanics [41], which are 'zero-consuming' in static states. Energy consumption can thus be significantly lowered. In the coming future, it is unlikely the complexity of physical ONN systems should surpass that of practical ANNs. The key to implement sophisticated ANNs, as demonstrated by this work, is temporal multiplexing with the assistance of electronic devices [10]. Thanks to recent progresses in hybrid integration of electronics and photonics [42, 43], it is optimistic to build a monolithic system with the OCDC for high-performance computing and affiliated electronics for instructions and memory.

As the prosperity of modern AI relies heavily on the success of deep learning regression, we demonstrate the OCDC to promote the application of ONNs in regression tasks. In our approach, firstly, values in the complete real domain are represented by optical fields. Output values are detected via optical coherent interference to maintain amplitude information. Secondly, the size of neural networks that are used in regression tasks is far larger than that of currently available ONNs. We reconfigure and reuse the OCDC so that matrix multiplications and convolutions of arbitrary size can be equivalently conducted. Thirdly, the OCDC features precise control through BPC method to reach a high numerical accuracy (normalized deviation of ~0.01). Therefore, the OCDC meets the prerequisite for sophisticated deep learning regression tasks. A state-of-art image reconstruction neural network, AUTOMAP, is adopted to benchmark the proposed chip. Experimental results validate that the OCDC can accomplish AUTOMAP with comparable performance of the 32-bit digital computer. Since the basic building blocks of ANNs across different applications are similar, we believe the OCDC can be further applied in more advanced AI fields, including autonomous driving, natural language processing, medical diagnosis, and scientific study.

## Materials and methods

**Experimental setup.** A brief schematic of the experimental setup is illustrated in Fig. 3d. A continuous-wave laser (Alnair Lab TLG-220) working at 1550 nm is used as the coherent light source. The output of the OCDC is directly linked with the amplifier photodetector (THORLABS PDA10CS2). The electrical signal is digitized and recorded with an oscilloscope (KEYSIGHT DSO-S 804 A). A multi-channel arbitrary waveform generator (AWG) is used as the signal source for the OCDC. The AWG contains 9 NI-PXIe-5413 blades, each of which has two output channels, and an embedded computer (NI-PXIe-8880) to control the system. The signals recorded by the oscilloscope is also transmitted to the computer for further processing. In the experiment, we use three branches (six modulators) to demonstrate the OCDC's feasibility on dot product. Since each modulator requires 2 electrical signals for push-pull driving, 12 output channels of the AWG are adopted. The trigger of the oscilloscope and the AWG is synchronized for stable sampling. A homemade 45-channel voltage source (VS) is used for supplying the bias voltages of the modulators. The control signal is also provided by the embedded computer. As the proof-of-concept, single ended photodetection is used in this work, so that the photocurrent is quadratic to the optical amplitude. To achieve a linear mapping between electrical signal and optical amplitude, homodyne photodetection is preferred (see Fig. S8b). Note that correct operation of OCDC is based on stable coherent interference, a thermo-resistor and a thermo-electric cooler are packaged in the module and are controlled by a temperature controller (Thorlabs ITC4002QCL). The module is stored and measured with indoor humidity of 30% to 70%. Long-time exposure to high humidity might shift the position of fiber port, increasing the insertion loss. Therefore, it is preferred to use airtight seals in module packaging.

**Push-pull modulation.** In the OCDC architecture, the final output is yielded by constructive interference. It is important to keep the phases of different branches stable to maintain the constructive interference. Ordinary single-driven configuration of modulators not only impose amplitude modulation, but also introduce extra phase shift. Therefore, the push-pull configuration is necessary for the OCDC. In the experiment, thermo-optic modulators are adopted, where phase shift from the thermal effect is proportional to applied power (quadratic to voltage). Therefore, the relation between applied voltages and the expected phase difference $\Delta\varphi$ is formulated as

$$V_{upper} = \sqrt{\frac{P_0 + \frac{\Delta\varphi}{2\pi} \cdot P_\pi}{R}} \qquad V_{lower} = \sqrt{\frac{P_0 - \frac{\Delta\varphi}{2\pi} \cdot P_\pi}{R}} \tag{2}$$

where $V_{upper}$ and $V_{lower}$ is the voltage applied to the upper arm and the lower arm of a modulator, respectively. $P_0$ is a bias power applied to the upper and lower arms at the same time. This bias power allows negative phase shift to the lower arm, thus realize push-pull configuration. $P_\pi$ is the required power for 180° phase shift, which is different for every single thermal phase shifter. $R$ is the resistance of the thermal phase shifter, which is measured to be 1.6 kΩ. For simplicity, for all thermal phase shifters, the $P_0$ is set at 2.28 mW, and an approximate value of $P_\pi$, 1.81 mW, is used. With a proper bias voltage, the transmission function of the modulator is $sin(\Delta\varphi)$. With this transmission function, we encode the pixel values in grey scale to voltages for modulation. We have also built an electro-optic (EO) version of OCDC (see Suppl. for more information) with silicon-based modulators. To achieve push-pull modulation on these EO modulators, a common positive bias voltage should be applied to the phase shifters of upper arm and lower arm, before the differential voltage signals are applied.

**Architecture of the AUTOMAP.** The AUTOMAP is composed with, from first to last, two fully-connected (FC) layers, two convolutional layers, and a de-convolutional layer. The input images are flattened to an input vector of the first FC layer. For MF and vPDS processes, the shape of input images is 128×128×2. Therefore, the length of input vector is 32,768. For Radon process, the size of input image is 185×180=33,300. The number of output neurons in the first layer and the second layer is 8,100 and 16,384, respectively. The original number of neurons in the first layer is 16,384. In our experiment, this number is modified due to memory limitation of our training platform (NVidia RTX-2080ti with 10.6 GB graphic memory). The activation function of FC layers is 'tanh'. After two FC layers, the resultant vector is transformed back to the form of an image with the shape of 128×128 for 2D convolutions. The number of output feature maps of these two convolutional layers is 64, and their size is 128×128. The kernel size is 5×5, and the activation function is ReLU. The kernel size of the de-convolutional layer is 7×7. The de-convolutional layer directly outputs the image result without activation.

**Neural network training.** The dataset for image reconstruction is assembled from four subjects of the MGH-USC HCP program [44]. 3100 sagittal scanned brain magnet resonance images (MRI) are used for neural network training and validation. The original size of these images is 256×256, and they are undersampled to 128×128 for the capability of the AUTOMAP. The maximal values of these images are normalized to 1 and the mean values are subtracted.

The AUTOMAP is a generally feasible network that can reconstruct images from various processes with the same network hyperparameters. In this work, we demonstrate 3 processes: reconstructing images from misaligned Fourier (MF) spaces [35], from undersampled Fourier spaces [36], and from Radon projections [37].

- The MF process. In MRI, ghost images often occurs when the Fourier spaces of two trajectories are physically misaligned. The AUTOMAP is trained to reconstruct images without ghost from the misaligned Fourier spaces. For training, misaligned Fourier spaces are generated from the original images. The images are firstly transformed to their Fourier space by fast Fourier transform (FFT). Then an extra phase shift is added to the even row of the Fourier space. The real part and the imaginary part of the processed Fourier spaces are used as the inputs of the AUTOMAP and the original images are the ground truths for training.
- The undersampled Fourier process. In this process, the AUTOMAP is trained to reconstruct images from the Fourier spaces which have been sparsely undersampled, i.e., only a few pixels are reserved and others are set to zeros. To generate these undersampled Fourier spaces, we adopt variable Poisson disk sampling (vPDS) method with a sparsity of 0.6 to the Fourier spaces transformed from the original images. Again, during training, the real part and the imaginary part of the undersampled Fourier spaces are used as inputs and the original images are used as the ground truths.
- The Radon process. Radon projection is a conventional method for MRI. Here, the AUTOMAP is trained to perform inverse Radon transform with better quality than the conventional one. The Radon projection is generated directly from the original images using discrete Radon transform (180 projection angles with 185 parallel rays). These Radon projections are used as inputs and the original images are the ground truths for training.

The dataset with 3100 examples is randomly divided into a training set (2700 examples) and a validation set (400 examples). The loss function of training is formulated as

$$L^{\Theta} = \frac{1}{N^2}\sum_{i=1}^{N}\sum_{j=1}^{N}\left(y_{i,j}^{\Theta} - \hat{y}_{i,j}\right)^2 + \frac{\lambda}{KN^2}\sum_{s=1}^{K}\sum_{i=1}^{N}\sum_{j=1}^{N}\left|h_{s,i,j}^{\Theta}\right| \tag{3}$$

where $y^{\Theta}$ is the output image by the neural network, and $\hat{y}$ is the ground truth image. $h^{\Theta}$ represents the output feature maps of the second convolutional layer. $N$ = 128 is the width of the images, and $K$ = 64 is the number of feature maps. Penalty factor $\lambda$ is 0.0001. The optimization method is the 'Adam' optimizer [45] with learning rate of 2e-5. After 850 epochs of training, the learning rate is decayed to 2e-6 for better convergence. For different reconstruction processes, the neural network is trained independently. The training platform comprises an Intel Xeon-E5-2640v4 CPU and an Nvidia RTX-2080ti GPU, and the training time is 2h10min.

Figure S6 illustrates the loss functions during training (please see the 'CBD' curves). The training loss converges well and the validation loss does not show overfitting, indicating a successful training. As comparisons, we also show the performance of training when the AUTOMAP is not real-valued (see Fig. S6, Fig. S7, and Suppl. for further information). We find that if an ONN fails represent the complete real-value domain, it will perform poorly or even fail to reconstruct images. Therefore, the ability of amplitude modulation and coherent detection of the OCDC is necessary.

***In-situ* backpropagation control.** The OCDC is an analog computing hardware. The imperfections of the fabricated devices have a large impact on the final numerical accuracy. Even if we can calibrate the OCDC by measuring the modulation curve (transparency vs. applied voltage) of every modulator, the computing results still deviate from the desired ones when all modulators work simultaneously. We use backpropagation control (BPC) method to further minimize the deviations of the hardware. In contrast to previous ONN *in-situ* training method such as [46], our BPC is used to fine-tune parameters from a computer-pretrained network. Instead of updating the parameter as a whole, the BPC updates parameters independently and it is suitable for the OCDC to reach higher numerical accuracy. Assume $k \times k$ parameters are to be fine-tuned. The computing complexity of BPC is $O(k^2)$, which is at the same order of magnitude with the *in-situ* training method [46]. For a temporally multiplexed OCDC, the forward propagation is formulated as

$$y_i = \sum_{j=1}^{M} x_{i,j} \cdot w_j, i = 1, 2, ..., N \tag{4}$$

where $M$ is the number of branches used for dot product. It is 3 in the experiment. The time step $N$ is 250. By defining the mean square error (MSR) of the results as the loss function, we can calculate the derivatives of loss function ($L$) on the hardware-represented weights ($w$).

$$\frac{\partial L}{\partial w_j} = \frac{2}{N} \sum_{i=1}^{N} \left( y_i - \hat{y}_i \right) \cdot x_{i,j} \tag{5}$$

where $\hat{y}$ is the result of the desired dot product. Update these weights by changing the applied voltages on the modulator, the MSR of the results is minimized (more examples of BPC are illustrated in Fig. S10). Since the BPC is conducted on a digital computer, we estimate the overhead of this process. By assuming a single-core CPU working at 4-GHz clock speed, the backpropagation theoretically takes only 0.18 μs to finish. Considering the accuracy improvement provided by the backpropagation, such overhead is acceptable. Note that the forward propagation and backpropagation of the OCDC are all linear. Nonlinear distortions cannot be eliminated by this method. Therefore, such nonlinearity imposes a limitation for BPC (further discussed in the Suppl.).

**Conducting the AUTOMAP by the OCDC.** The linear part of the first FC layer and the first convolutional layer of the AUTOMAP is experimentally conducted by the OCDC chip. In the first FC layer, the size of input vector is 1×32768 and the size of the weighting matrix is 32768×8100. They are decomposed to small parts with the size of 1×3 and 3×8100, respectively. The values from the input vector are loaded to the 'slow mod.'

modulators and the values from the matrix is loaded to the ‘fast mod.’ modulators. Because of the massive weight matrix of the FC layer, it is impractical to conduct all operations with the OCDC working at a modulation rate of 100 Hz. The OCDC carries out operations for three typical parts in the input vector: the corner, the center, and the edge (see Fig. S11 for further information). For the convolutional layer, we conduct the convolution with the first kernel (5×5). The kernel is flattened to a vector and the input feature map is rearranged to a matrix with the size of 25×16384. Then, the decomposition method is similar to that of the FC layer. The OCDC conducts linear parts, and the nonlinear activation functions are implemented in the computer. From all the experimental results, we calculate the normalized standard deviation of residual errors that introduced by the OCDC. It is averagely 0.0076 for the FC layer and 0.0104 for the convolutional layer. In the image reconstruction processes, we impose these experimental deviations to every layer of the AUTOMAP as additive noise to simulate the situation that the neural network is completely conducted by the OCDC. Corresponding results are provided in Fig. 5 and Fig. S12-14. Additionally, the quality of image reconstruction with different levels of computing error is simulated and discussed in Suppl.

## Acknowledgement

This work is supported in part by the National Key Research and Development Program of China (Program No. 2019YFB2203700) and the National Natural Science Foundation of China (Grant No. 61822508).

## Conflict of interests

The authors declare no competing interests.

## Author contribution

W. Z. initiated this project. W. Z., X. W., and J. L. supervised the research. S. X. conceived the research and methods. J. W., H. S., B. B., and Z. Z. contributed in chip designing, fabrication, and packaging. S. X. and S. Y. built up the experimental setup and conducted the experiment. All authors contributed in the discussion of the research. S. X. prepared the manuscript with inputs from H. S., B. B., and Z. Z.

## Data availability

The brain images used for training and evaluation were obtained from the MGH-USC HCP database (https://db.humanconnectome.org/).

## Code availability

Source code is available from the corresponding author upon reasonable request.

# Supplementary information for Optical coherent dot-product chip for sophisticated deep learning regression

Shaofu Xu[1], Jing Wang[1], Haowen Shu[2], Zhike Zhang[3], Sicheng Yi[1], Bowen Bai[2], Xingjun Wang[2], Jianguo Liu[3], Weiwen Zou[1]*

*[1]State Key Laboratory of Advanced Optical Communication Systems and Networks, Intelligent Microwave Lightwave Integration Innovation Center (imLic), Department of Electronic Engineering, Shanghai Jiao Tong University, 800 Dongchuan Road, Shanghai 200240, China*

*[2]State Key Laboratory of Advanced Optical Communications System and Networks, Department of Electronics, School of Electronics engineering and Computer Science, Peking University, Beijing 100871, China*

*[3]Institution of Semiconductors, Chinese Academy of Sciences, Beijing 100083, China*

**Correspondence to: wzou@sjtu.edu.cn.*

## S1. Chip fabrication and characterization

The on-chip devices in this paper are designed and fabricated on the silicon on insulator (SOI) platform with 0.13 μm mask technology. Besides the thermo-optic version of OCDC used in the proof-of-concept experiment, an electro-optic (EO) version of OCDC, which is consistent with the architecture, has also been fabricated and measured for future demonstration in high-speed scenarios. As shown in Fig. S1a, the silicon Mach-Zehnder modulator here works under the carrier-depletion mode which ensures the 3 dB EO bandwidth of above 24 GHz, as indicated in Fig. S1b. Therefore, it is promising to improve the computing speed for about several gigahertz. The insertion loss of each modulator is ca. 5dB.

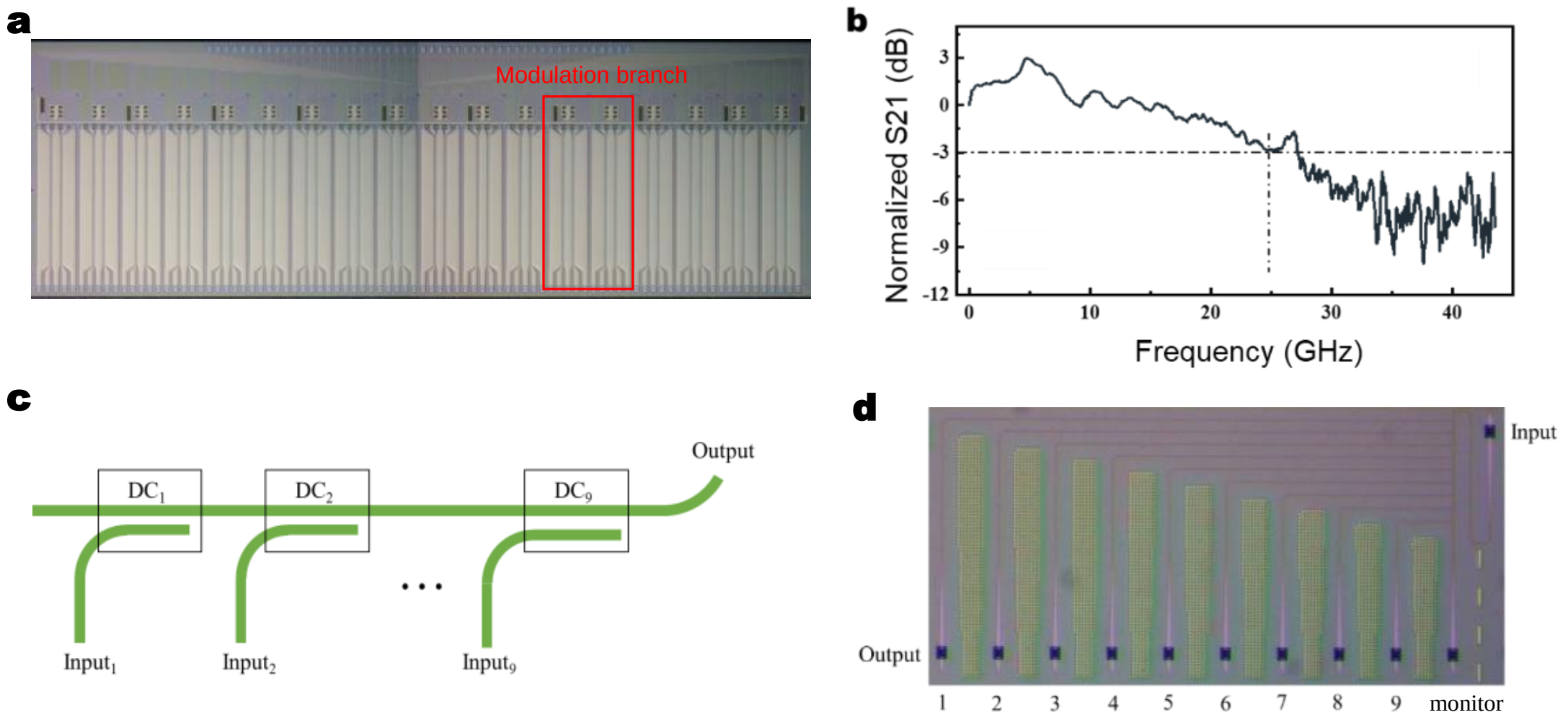

**Fig. S1 a,** The top view of the EO version of the OCDC. A modulation branch is marked with a red block, including two cascaded electro-optic modulators and a tail phase shifter. **b,** Bandwidth measurement of fabricated EO modulators. The 3-dB bandwidth is over 24 GHz. **c, d,** the schematic illustration (**c**) and optical image (**d**) of the 1x9 cascaded directional coupler (DC) array.

The dot-product process consists of both multiplying operation within one modulation branch and adding operation of all branches, which may result in calculation deviations if uneven power splitting happens. A cascaded directional coupler (DC) array is introduced to achieve even power splitting. The schematic is shown in Fig. S1c. Ideally, each branch should obtain 1/9 of the input power, and that will make the power splitting ratio distribution $1/i$, where $i$ is the port number. Moreover, a residual power level of $P_r$=~30% is allocated to the monitor port. In that way, the actual coupling strength of each DC can be described as $P_r/i$. In our design, the width of the silicon bus waveguide of the cascaded DC structure is 450 nm and the gaps between bus waveguide and each branch are set to be 250 nm, with the coupling length varying from 5.2 μm to 8.1 μm. Figure 2d in the main text shows the test results of the normalized power at each output port. One can tell that the power deviation is less than 1.2 dB among all ports, showing a considerable result with uniformed power splitting.

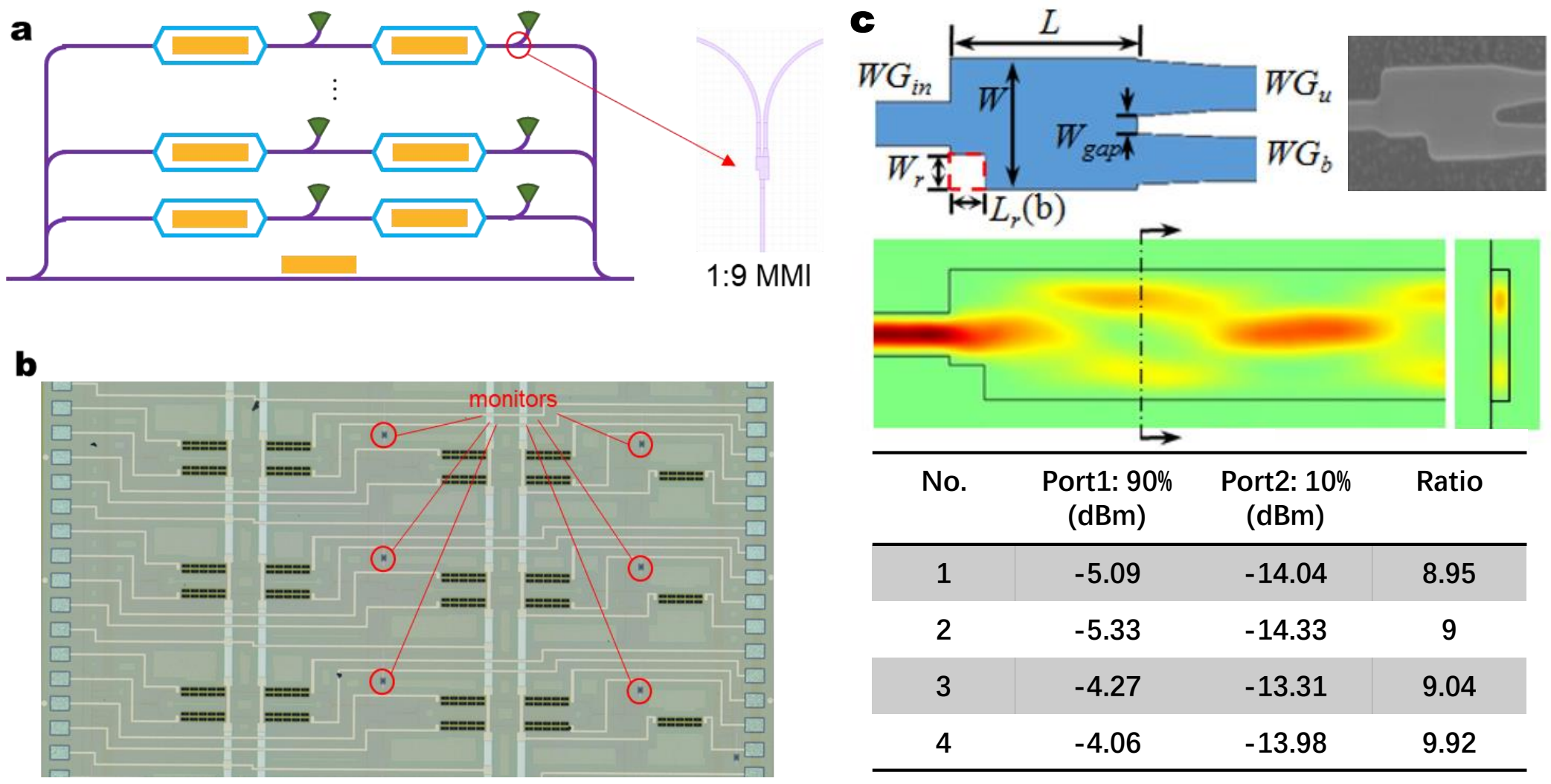

| No. | Port1: 90% (dBm) | Port2: 10% (dBm) | Ratio |
|---|---|---|---|
| 1 | -5.09 | -14.04 | 8.95 |
| 2 | -5.33 | -14.33 | 9 |
| 3 | -4.27 | -13.31 | 9.04 |
| 4 | -4.06 | -13.98 | 9.92 |

**Fig. S2** Deployment of the monitors on the chip. **a,** Schematic of the OCDC, where the monitors are marked with green triangles. The optical power for monitor is divided from the output port of each modulator via a 1:9 MMI coupler. **b,** The top view of three modulation branches on the thermo-optic OCDC, showing six monitors on the chip. **c,** Design, fabrication and measurement results of the 1:9 MMI.

We deploy a coupler at the output of every modulator so that we can monitor and characterize the performance of all modulators individually. As shown in Fig. S2, optical power for monitoring is divided from the output of the modulator using a 1:9 asymmetric multimode interferometers (MMIs). Such splitting ratio is constructed by simply breaking the symmetry of the multimode region [1]. The proposed asymmetric MMI power splitter is shown in Fig. S2c. Compared to the conventional symmetric power splitter, the only difference is that the symmetry of the multimode region is broken by removing its bottom left corner (marked with a red dashed rectangle). Such a minor structural change causes a dramatic redistribution of the optical field thus leading to an uneven power splitting by changing the value of $L_r$. We randomly chose four identical 1:9 MMI and test the power splitting ratio. The results are found to be very close to the design target (9.0), as indicated by the bottom right table in Fig. S2c.

Using these monitors, we measure the maximal optical power of every modulator by changing the bias voltage to their maximal transparency. Results are shown in the table of Fig. S3. The imbalance of the 9 branches in the thermal-optic OCDC is around 2.6 dB and it is around 3.8 dB for the electro-optic OCDC. The difference of optical power between the thermo-optic OCDC and the electro-optic OCDC comes from the deviation of fiber edge coupling at the light input port.

**a**

| Branch number | Optical power (dB) | |
|---|---|---|
| | Stage 1 | Stage 2 |
| 1 | -32.4 | -33.5 |
| 2 | -31.8 | -33.1 |
| 3 | -34.2 | -35.5 |
| 4 | -33.2 | -34.3 |
| 5 | -33.8 | -34.8 |
| 6 | -31.6 | -32.9 |
| 7 | -33.1 | -34.4 |
| 8 | -32.6 | -33.9 |
| 9 | -32.8 | -34.0 |

**b**

| Branch number | Optical power (dB) | |
|---|---|---|
| | Stage 1 | Stage 2 |
| 1 | -24.2 | -28.6 |
| 2 | -25.0 | -30.8 |
| 3 | -26.5 | -31.8 |
| 4 | -25.8 | -30.9 |
| 5 | -25.0 | -28.0 |
| 6 | -26.8 | -31.0 |
| 7 | -25.6 | -29.5 |
| 8 | -26.5 | -31.6 |
| 9 | -26.8 | -31.5 |

**Fig. S3** Tables of measured optical power after every modulator on the thermo-optic OCDC (**a**) and the electro-optic OCDC (**b**), respectively. It is found that the uniformity and insertion loss of the thermo-optic modulators is superior than the electro-optic ones.

The method of module packaging is illustrated in Fig. S4 and Fig. S5. For the OCDC, there are 36 signal pads and the space of the pads is 250 μm. In consideration of the ease of usage and highly configurable feature, the high-speed socket with up to 100 positions (Samtec, ERF8-RA) is selected as the optimal solution for the compact module, as shown in Fig. S4. The 60 DC pads have been connected using a flexible printed circuit (FPC) connector. In order to match the silicon-based waveguide mode field and increase the coupling efficiency, the tapered fiber with ~3.5 μm mode field is used for optical coupling. For RF transmission path configuration, a transmission line printed circuit board (PCB) with preferable mechanical strength is required to place high speed socket. And a transmission line AlN circuit board (ACB) is employed to connect the OCDC and the transmit line PCB. In order to ensure the high frequency performance, the impedance matching of the transmission line is required. The high frequency performance of the RF transmission path is measured as shown in Fig. S5. Insertion shows the measurement scheme in which the pinboard with GPPO connectors and the high frequency probe are used. It can be seen that the -3-dB bandwidth can all reach up to 7.8 GHz for the center line and the edge line, respectively. In this sense, the intrinsic performance of the OCDC is ensured.

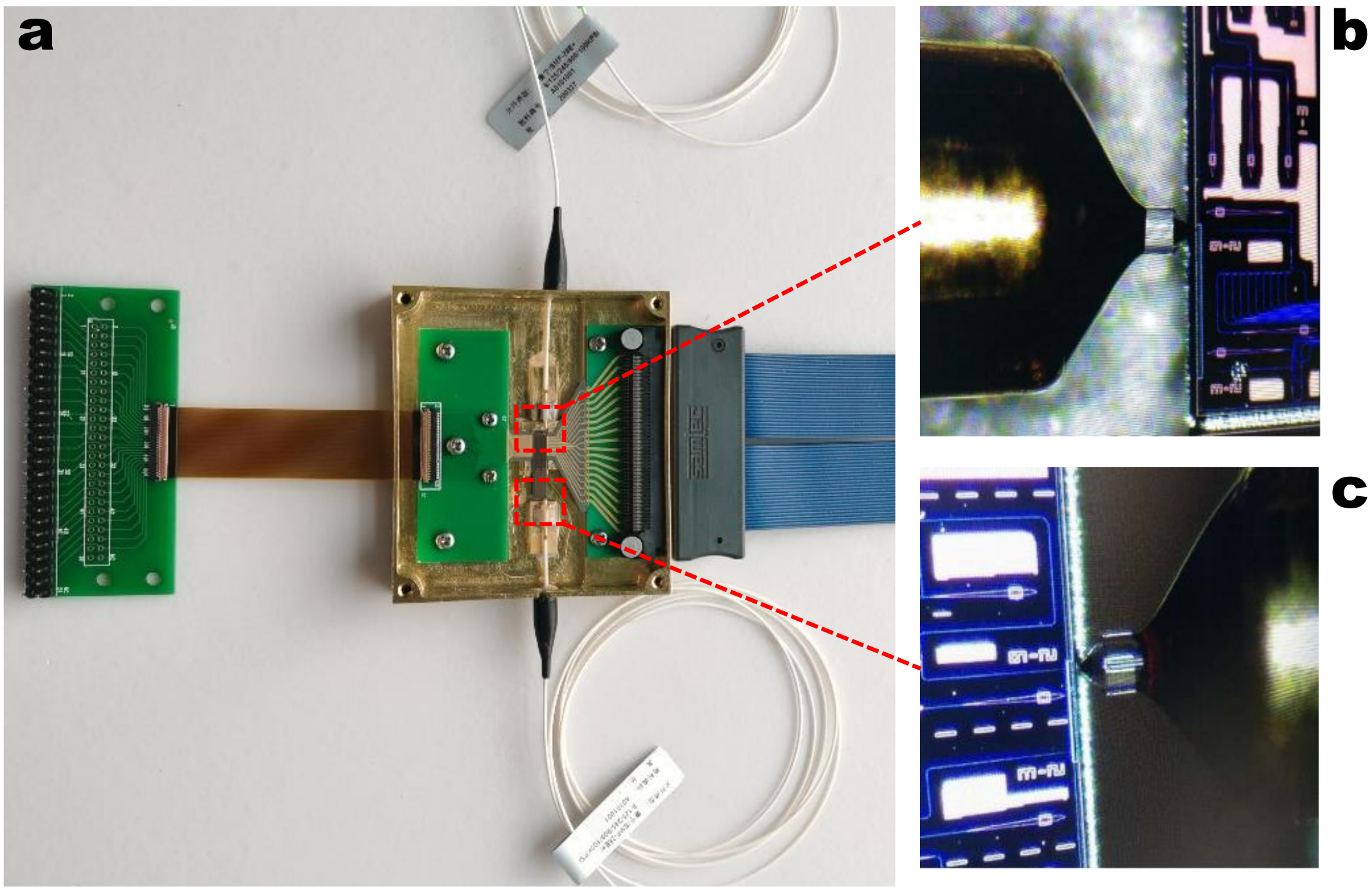

**Fig. S4** The packaged OCDC with measurement setup. **a**, the photograph of packaged module in which the DC and RF measurement device is shown. **b** and **c** are the input and output coupling tapered fiber of the OCDC, respectively.

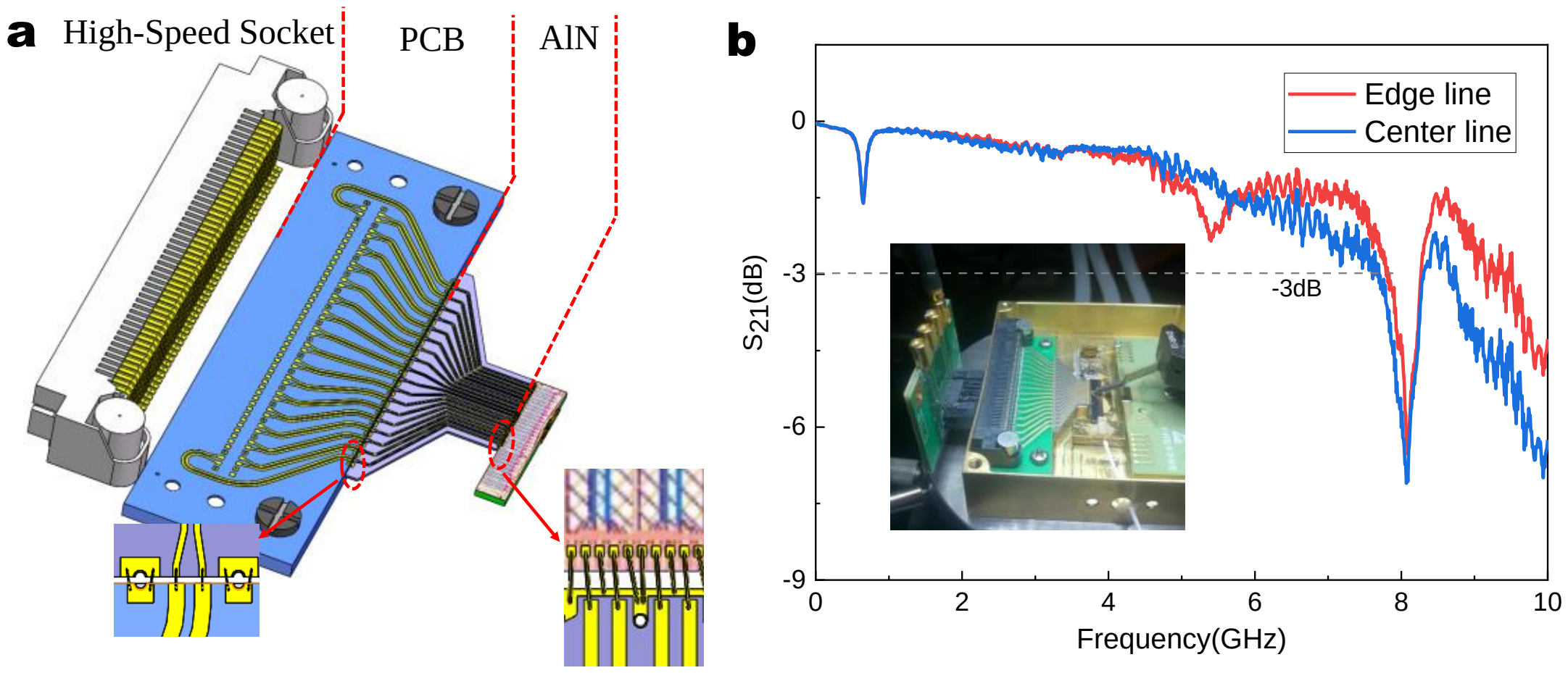

**Fig. S5** The RF transmission characteristics of the module. **a**, the RF transmission path configuration including a high-speed socket, a transmit line PCB and a transmission line ACB. **b**, The frequency response of the whole RF transmission path. For the edge line and the center line, the -3-dB bandwidth can reach up to 7.8 GHz which can support the performance of the OCDC. Insertion is the measurement scheme.

## S2. Performance of the AUTOMAP in incomplete real-value domains

The numerical basis for most regression tasks is the complete real-value domain, as shown in Fig. 1. However, the capability of practical ONNs of realizing complete real-value domain may be limited by the physical constraints. For example, architectures using non-coherent light to represent values are not able to conduct operations with negative input values. With balanced detection, weights and outputs can be real-valued. Similarly, coherent architectures use optical amplitude to represent real-valued inputs and weights. However, using single-ended photodetection is unable to output negative values. Besides, architectures that rely completely on intensity modulation, intensity attenuation, and intensity detection can only conduct operations with non-negative inputs, weights, and outputs. The numerical domain of the above situations is not complete. Different from these architectures, the OCDC proposed in this work can represent real-valued inputs and weights with amplitude modulators. It can also output real values because the output amplitude is biased by optical interference prior to detection.

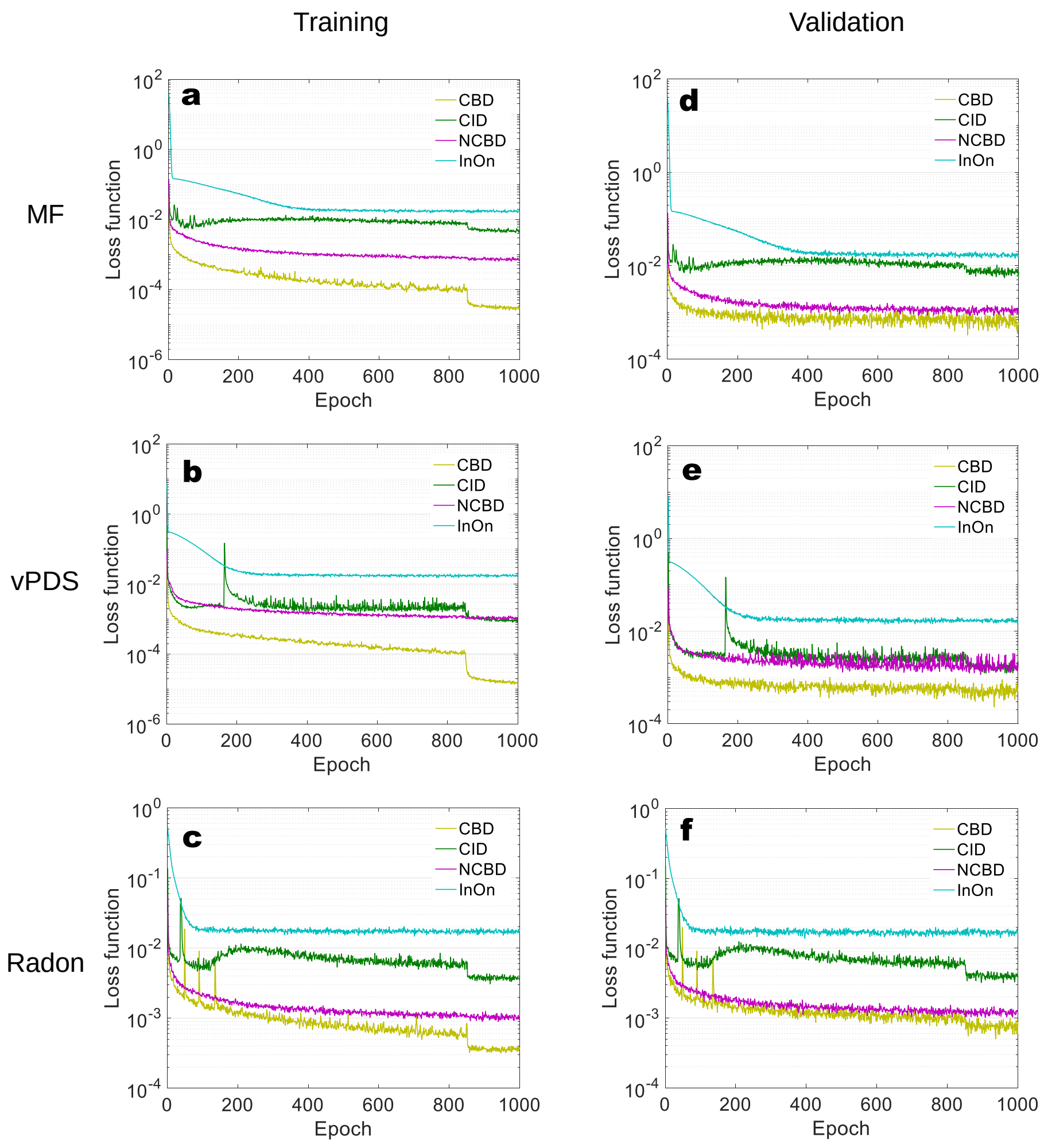

**Fig. S6** Loss functions of the AUTOMAP during training. **a-c**, Training losses of different architectures on different reconstruction processes. Note that the training loss implies fitting ability of a neural network. The fact that the CBD architecture outperform other architectures indicates the necessity of the complete real-value domain for deep regression tasks. **d-f**, Validation losses, which implies the generalizability of a neural network.

In this section, we investigate the performance of the AUTOMAP if the real-value domain is not complete. As discussed above, there are four situations: coherent architectures with single ended intensity detection (CID), non-coherent architectures with balanced detection (NCBD), architectures with intensity-only capability (InOn), and coherent architecture with biased detection (CBD, this work). The AUTOMAP is modified to simulate the incompleteness of the real-value domain in different situations except for the CBD situation. To simulate the performance of CID architectures, absolute value operation is imposed to the results of linear part of each layer. For NCBD architectures, the activation function of FC layers is replaced with ReLU to give non-negative input values. For InOn architectures, absolute value operation is imposed to the weights and output results of each layer. Every modified network is trained independently on three reconstruction processes (MF, vPDS, and Radon).

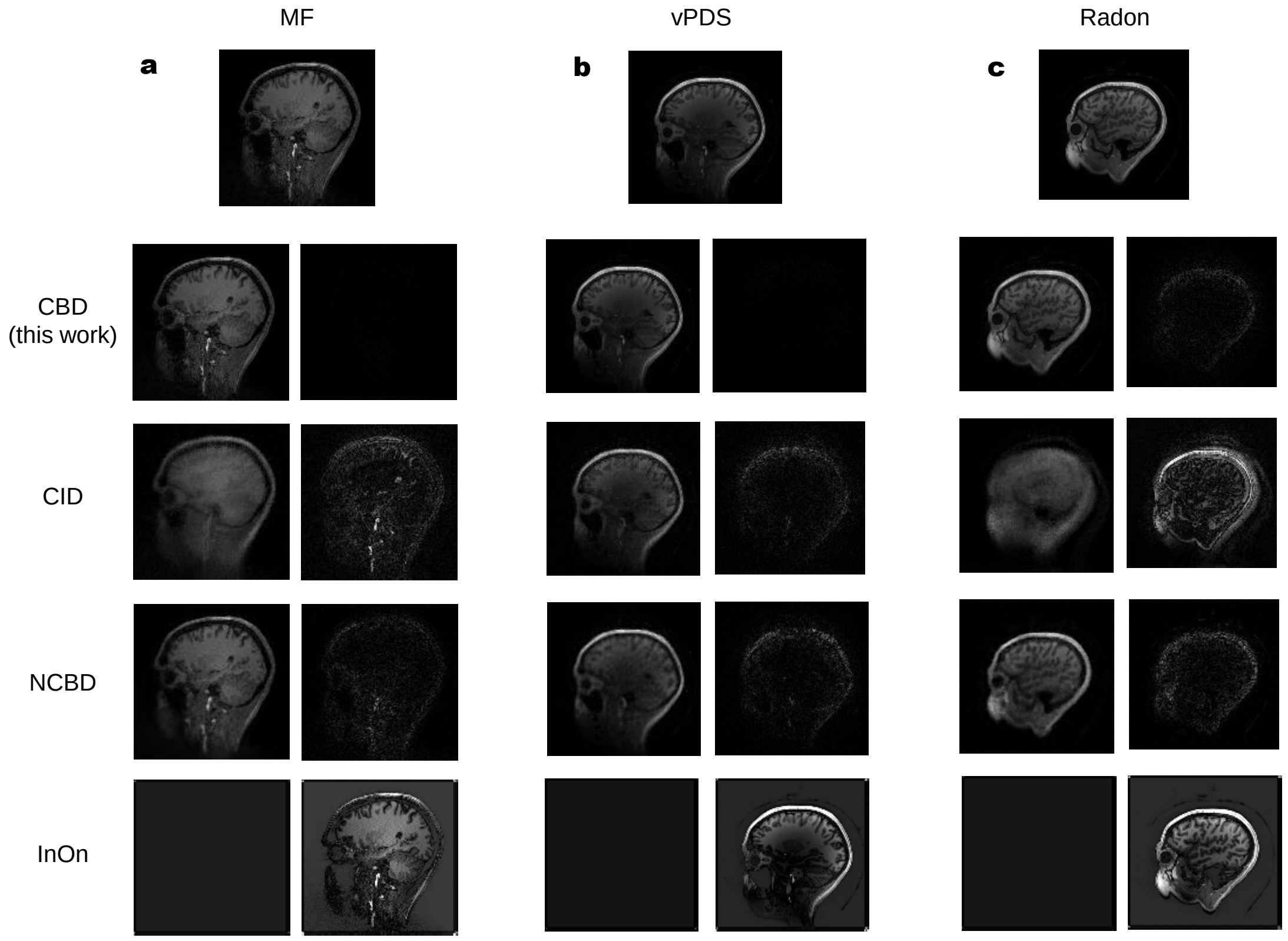

**Fig. S7** Reconstructed images by these four architectures (CBD, CID, NCBD, and InOn). **a-c**, The MF

process, the vPDS process, and the Radon process, respectively. The ground truth image is offer on the top. On the right side of each reconstructed image, residual error image is attached (values are amplified by 2 times for visibility).

Results are shown in Fig. S6 and Fig. S7. From the loss functions of training and validation, we observe that the CBD architecture can obtain the best convergence among these four situations. The NCBD and CID architectures can also converge but with inferior performance. The InOn architecture is unable to converge. Such difference in performance is obviously shown in Fig. S7. If the real-value domain is incomplete, the quality of image reconstruction will be much inferior. The InOn architecture even fails to conduct image reconstructions.

## S3. Calibration of the OCDC

The first step of calibration is to control the bias voltages of every modulator. The modulator is push-pull configured and the input signal is a saw-like wave. If the modulator is set to the null point, the output waveform should be a standard sine wave. The drift of bias voltage leads to harmonic distortion of even orders. We calibrate the bias voltages such that the second-order harmonic distortion of the output waveform reaches the minimal.

The second step is to obtain the constructive inference among different branches. The optical combiner used in the OCDC is cascaded directional couplers (DCs), which is symmetrically the same as the optical splitter. As shown in Fig. S1c, the combining ratio is 1:1, 1:2, …, 1:9, respectively. Optical interference is occurring at every DC, which can be formulated as

$$A'_{n+1} = \begin{bmatrix} \sqrt{\dfrac{n}{n+1}} & \sqrt{\dfrac{1}{n+1}} \end{bmatrix} \begin{bmatrix} A'_n \\ A_n \end{bmatrix}. \tag{s1}$$

Eq. s1 describes the interference of the $n$-th DC. $A_n$ is the complex amplitude of $n$-th input port. The output field of interference $A'_{n+1}$ is the input of the next DC. $A'_1 = A_0$. After 9 DCs, the final output amplitude is formulated as

$$A_{out} = \frac{1}{\sqrt{10}} \sum_{n=0}^{9} A_n \tag{s2}$$

It is found that the cascaded DCs perform equivalently as a single ten-port optical combiner with uniform combining ratios. Eq. s2 indicates that we can calibrate the phases of all branches by measuring the final output field instead of the output field of every DCs. With this conclusion, it is possible to observe the in-phase condition of all branches. With the modulators tuned to maximal transparency, the final output field reaches the maximum when all branches are in-phase.

## S4. Limitation of backpropagation control

In general, the performance of the backpropagation control (BPC) method is limited by two factors: noise and nonlinear effects. The influence of random noise can be eliminated by increasing the time step ($N$ in Eq. 4 and Eq. 5) used for BPC. Because the noise can cancel each other out at the averaging process. However, it is not recommended to use a very large time step. With increased $N$, The BPC will consume more time and more resources to calculate the gradients. Besides, although the BPC can be very precise, the numerical accuracy of forward propagation is still affected the system noise. The extra accuracy of BPC does not contribute to better performance of dot product.

The second factor of the performance limit is the nonlinearity introduced by the 'fast mod.' modulators. Since we assume that the OCDC performs dot product linearly, the BPC is only capable to compensate the linear deviations of the forward propagation. However, nonlinearity always exists, in practical modulators. Despite the inherent nonlinearity of phase shifter, in our experiment, the major part of nonlinearity comes from the imbalanced phase modulation, i.e. phase shifts of the upper arm and the lower arm are not equal. As stated in the Methods, we use an approximate $P_\pi$ for simplicity. Given that the actual $P_\pi$ for every thermal phase shifter is not the same, such approximation will introduce nonlinearity.

Suppose the accurate $P_\pi$ for upper arm and the lower arm are $P_\pi^{(u)}$ and $P_\pi^{(l)}$, respectively. The loaded power for both arms is $\pm P_0$. The output optical field of the push-pull modulator is formulated as

$$
\begin{aligned}
E &= \frac{1}{2}\left( e^{jP_0\pi\frac{1}{P_\pi^u}} + e^{jP_0\pi\frac{-1}{P_\pi^l}} \right) \\
&= \cos\left( P_0\pi\frac{1}{P_\pi^u} \right) + j\sin\left( P_0\pi\frac{1}{P_\pi^u} \right) + \cos\left( P_0\pi\frac{1}{P_\pi^l} \right) + j\sin\left( P_0\pi\frac{-1}{P_\pi^l} \right) \\
&= \cos\left( \frac{P_0\pi}{2}\left( \frac{1}{P_\pi^u} + \frac{1}{P_\pi^l} \right) \right) e^{j\frac{P_0\pi}{2}\left( \frac{1}{P_\pi^u} - \frac{1}{P_\pi^l} \right)} \\
&= \cos\left( \frac{P_0\pi}{2}\Sigma \right) e^{j\frac{P_0\pi}{2}\Delta}
\end{aligned} \tag{s3}
$$

We can find a residual phase shift besides the standard push-pull modulation (the cosine function). This residual phase shift term distorts the linearity when it is detected by a photodiode. Suppose the reference optical field for photodetection is $A_{ref}$. The output photocurrent is formulated as

$$
\begin{aligned}
I_{pho} &\propto \mathrm{Re}\left\{ \left( \cos\left( \frac{P_0\pi}{2}\Sigma \right) e^{j\frac{P_0\pi}{2}\Delta} + A_{ref} \right) \cdot \left( \cos\left( \frac{P_0\pi}{2}\Sigma \right) e^{-j\frac{P_0\pi}{2}\Delta} + A_{ref} \right) \right\} \\
&= A_{ref}^2 + \cos^2\left( \frac{P_0\pi}{2}\Sigma \right) + 2\cos\left( \frac{P_0\pi}{2}\Sigma \right)\cos\left( \frac{P_0\pi}{2}\Delta \right)
\end{aligned} \tag{s4}
$$

which is not a perfect square expression. The square root of Eq. s4 is not linear to $\cos\left(\frac{P_0\pi}{2}\Sigma\right)$. A method to minimize such nonlinearity is to lower the deviation of $P_\pi$ between upper and lower arms, i.e., the $\Delta$ in Eq. s4. This depends on further advancement of photonic integration technologies. Another solution is to measure the $P_\pi$ parameter for every phase shifter so that we can compensate for such deviations by changing the applied power, $P_0$. However, this introduces more challenges to the calibration process and the encoding process from pixel values to voltages. Again, we note that the forward propagation suffers from system noise. Over-precise BPC is not helpful for the overall performance of dot product computing. There should be a balanced point between resource consumption and performance.

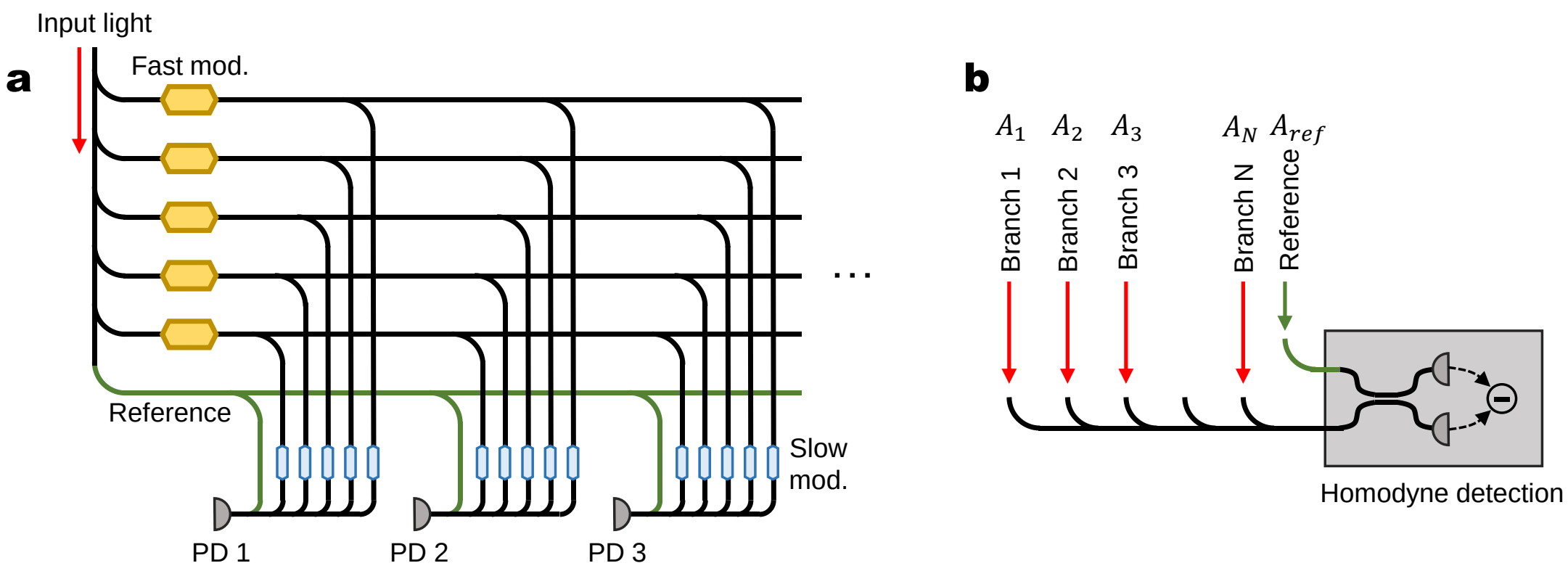

**Fig. S8 a,** The schematic of the spatially multiplexed OCDC. The input light splitter is the same as the OCDC presented in this work. After the 'fast mod.' modulators, the optical signals are again split into multiple 'slow mod.' banks. A bank carries out a dot product operation, and a PD outputs the result. **b,** The structure of homodyne detection. Using this structure, the output photocurrent is proportional to the amplitude of input optical field, e.g., $I_{pho} \propto A_{ref} \cdot \sum_i A_i \cdot \cos(\Delta\varphi)$.

## S5. Potential scalability of the OCDC

The basic concept of the OCDC is to reuse an optical dot-product core to perform matrix multiplications and convolutions of arbitrary size. While the temporal multiplexing of the OCDC is demonstrated in the main text, the OCDC can be also spatially multiplexed. The spatial multiplexing architecture of the OCDC is illustrated in Fig. S8a. the basic idea is to split the optical signals from the 'fast mod.' modulators to multiple 'slow mod.' modulators. As a single 'slow mod.' bank can conduct an operation of dot product, spatially multiplexing the OCDC can perform multiple dot product operations at once. In the architecture, the most energy-consuming and technically challenging part is the fast modulators. Therefore, the spatially multiplexed OCDC only duplicates the 'slow mod' part and photodetection part, to reuse the signals from fast modulators. The advantage of parallelism of optics can thus be exploited. Theoretically, with larger number of duplicates, energy efficiency of the OCDC becomes higher [2].

A key point of spatial multiplexing of the OCDC is low-loss waveguide crossings since they are largely adopted in the architecture. With current planar crossing technologies, insertion loss lower than 0.1 dB/crossing and crosstalk lower than 35 dB are obtainable [3], guaranteeing the implementation of large-scale spatially multiplexed OCDC. For more aggressive goals, multi-planar waveguide crossing can be adopted [4, 5]. The insertion loss and the cross talk can be further reduced. In the spatially multiplexed OCDC, the BPC is still feasible. Compared with the coherent ONN architecture based on Mach-Zehnder interferometer (MZI) mesh, BPC of the OCDC is straightforward and simple. In the MZI-based ONN, the deviation of each output port is influenced by every phase shifter in the mech. The whole weight matrix should be upgraded at the same time to minimize the deviation [6, 7]. However, in the OCDC, every output port is independent. One can easily determine which modulator introduces the deviation and calibrate them independently.

Another factor regarding the scalability is the computing density. The basic computing unit of the OCDC is an MZM (or MZI). Compared with microrings (MRRs), MZMs are much larger on footprint. So, the capability of conducting real-valued operation is at the cost of the computing density. According to the evaluation by T. Ferreira de Lima et al, MRRs can achieve a compute density about 50 TMAC $s^{-1}$ $mm^{-2}$ whereas MZI-based ONN can only achieve 0.56 TMAC $s^{-1}$ $mm^{-2}$.

## S6. The performance of AUTOMAP under different error level

Since the numerical accuracy of ONNs are highly relevant to the quality of deep learning regression. Here, we discuss how does the computing error influence the quality of image reconstruction. We generate different levels of random noises and add them into the ideal AUTOMAP to simulate that the AUTOMAP is carried out by the OCDC with different computing error. In this simulation, computing errors of the FC layers and the convolutional layer are the same. For every noise level, 200 images are reconstructed to show the stochastic result. The result is illustrated in Fig. S9.

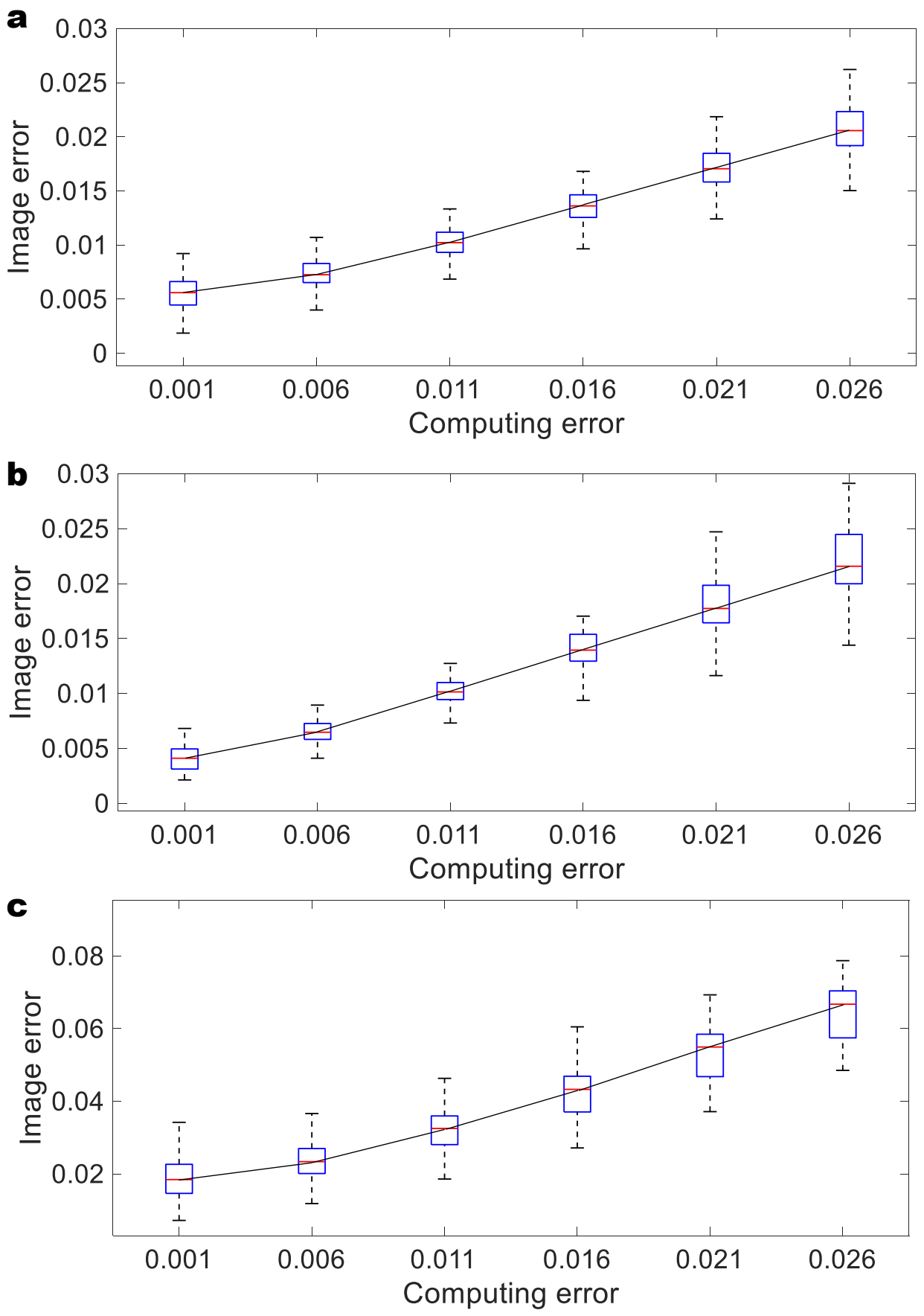

**Fig. S9** The error of reconstructed images with different levels of computing error. **a,** The Misalign reconstruction process. **b,** The vPDS reconstruction process. **c,** The Radon inverse transform reconstruction process.

It is observed that with higher computing error, the integral error of image reconstruction is higher. Except for the low-error region, the error of image reconstruction approximately grows linearly with the computing error. It is inferred that the quality of image reconstruction should be promoted linearly by achieving better numerical accuracy. When the numerical accuracy is high enough, the reconstruction quality is most governed by the capability of the neural network model.

## Supplementary references

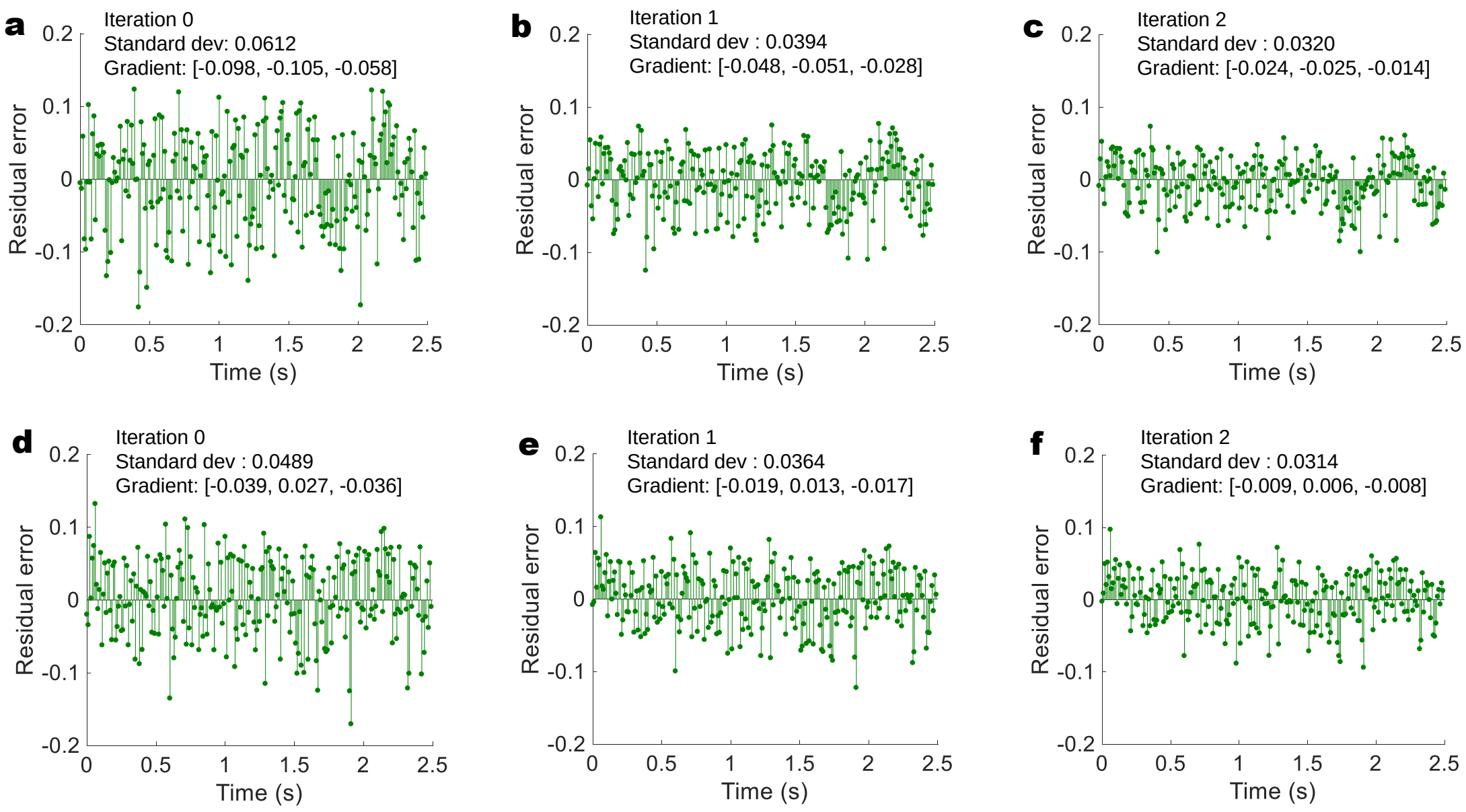

**Fig. S10** Residual error during BPC control. **a-c,** The adopted weights are [1, 1, 1]. Standard deviation of the residual error and corresponding gradients of the weights are shown in the figure. **d-f,** The adopted weights are [0.2, 1, 0.8].

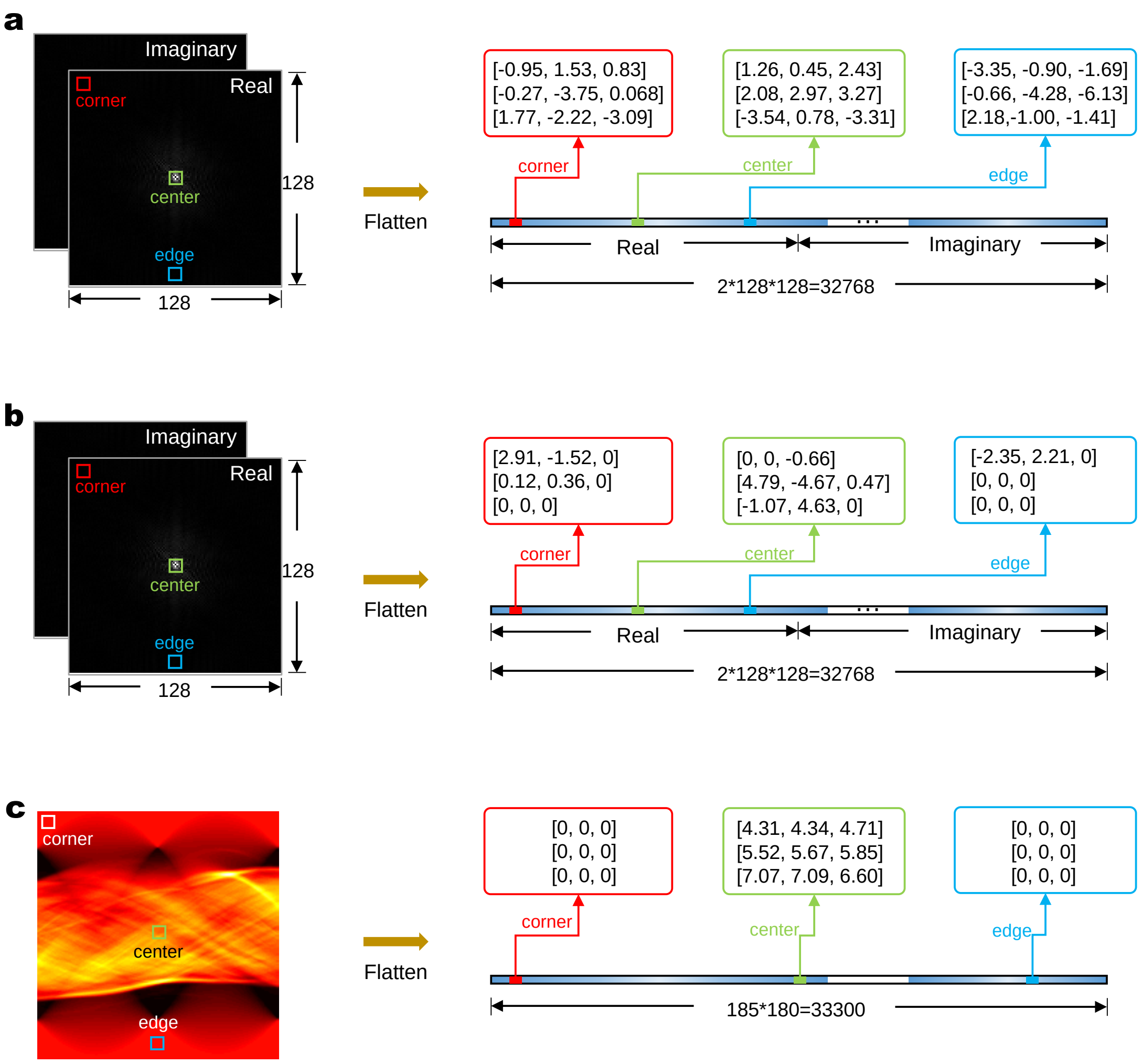

**Fig. S11** The input data preparation of the first FC layer. In the first FC of the AUTOMAP, the input images are flattened to a vector. For MF and vPDS processes, the input data includes a real image and an imaginary image. For Radon process, the input data is a real-valued sinogram. Sizes of these images are marked in the figure. Since the size of input vectors (32768) is significantly larger than the size of OCDC, it is extremely challenging to implement the complete matrix vector multiplication (32768×8100) experimentally with the modulation rate of 100 Hz. We used three typical parts (corner, center, and edge) on the image for OCDC calculation as a proof of concept. **a,** The data preparation for the MF process. After flattening, the pixels on the corner, the center and the edge of the real image are used for experimental implementation. The blocked values are the pixel values from these parts. **b,** The data preparation for vPDS is the same to that shown in **a**. **c,** The Radon process only has a real image. So, the locations of these three parts are different from those in **a** and **b**.

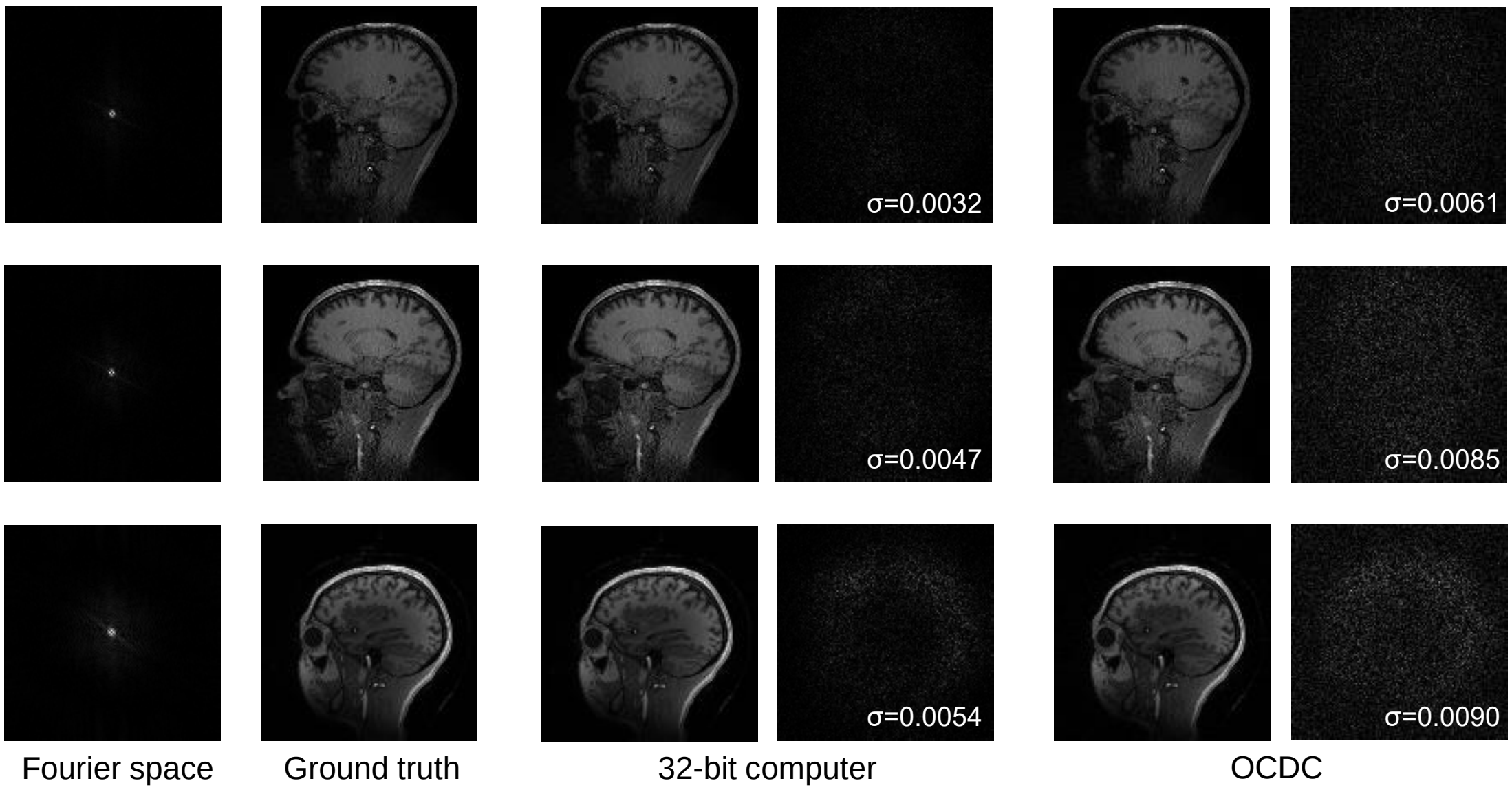

**Fig. S12** More examples of reconstructed images on MF process. Values of the error is amplified by 10 times.

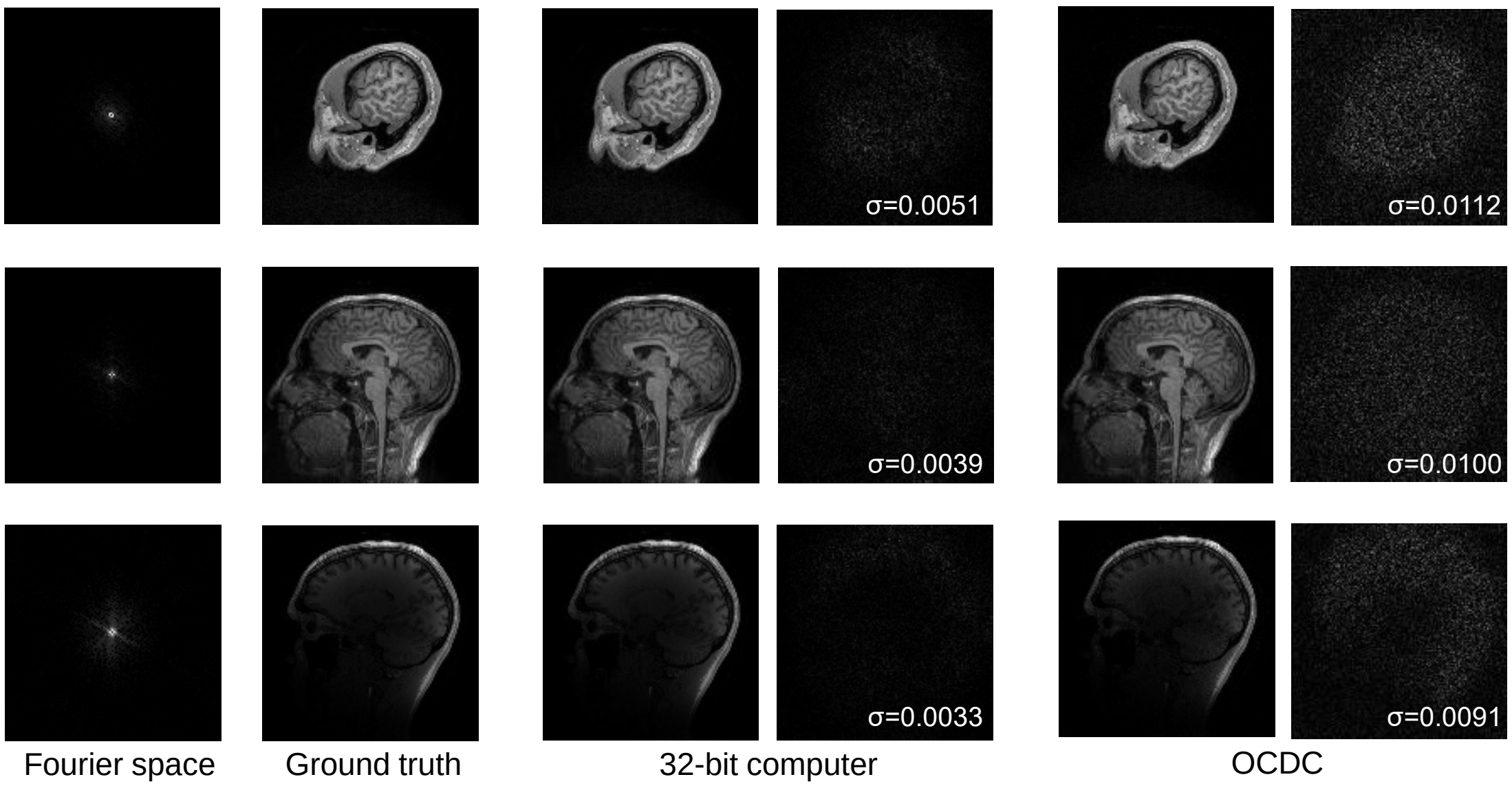

**Fig. S13** More examples of reconstructed images on vPDS process. Values of the error is amplified by 10 times.

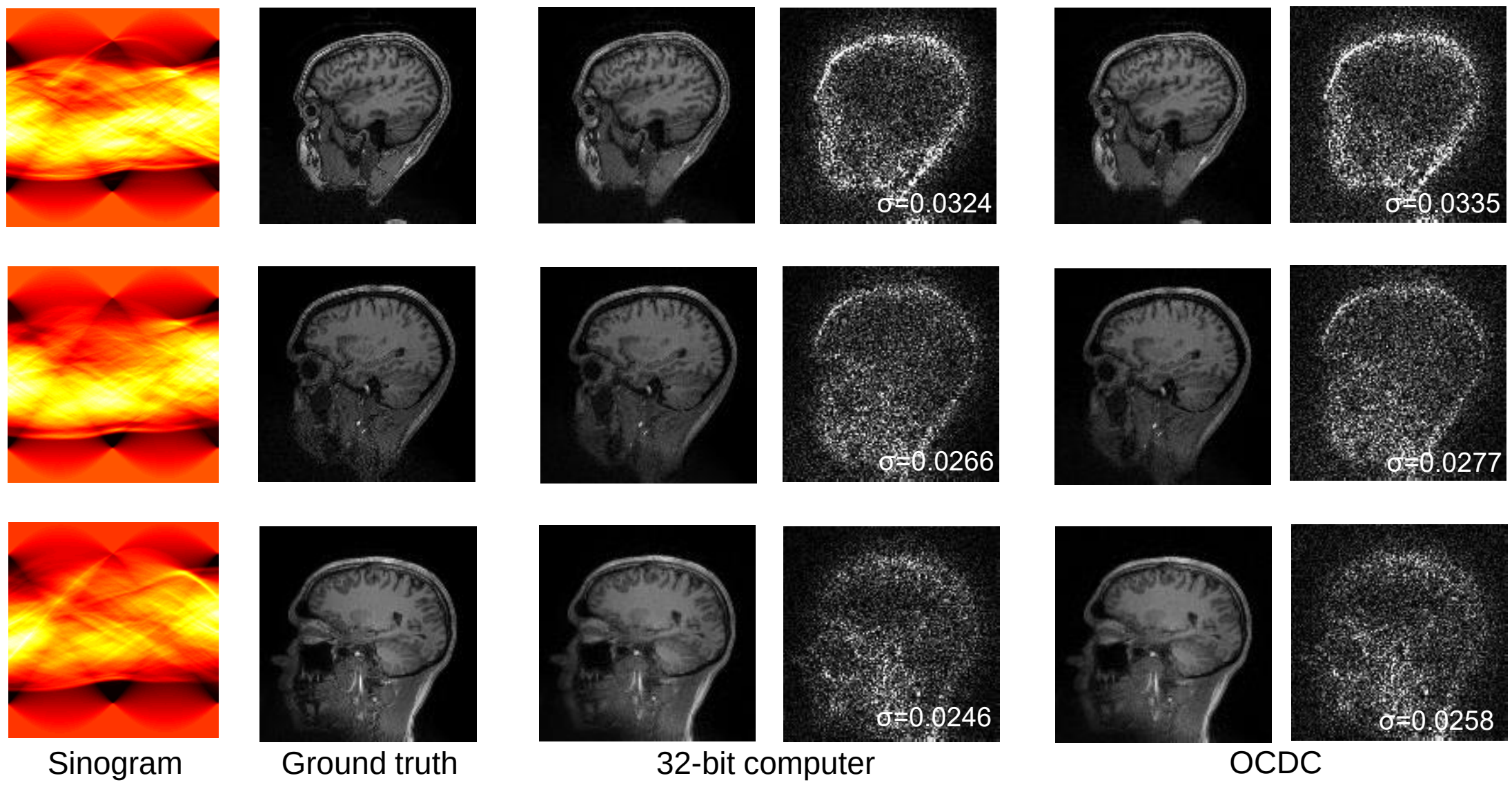

**Fig. S14** More examples of reconstructed images on Radon process. Values of the error is amplified by 10 times.